\documentstyle[12pt]{article}
\title{CONFINEMENT}
\author{Yu.A.Simonov\\ Institute
of Theoretical and Experimental Physics\\ 117259, Moscow,
B.Cheremushkinskaya 25, Russia}

\newcommand{\be}{\begin{equation}}
\newcommand{\ee}{\end{equation}}
\begin{document}
\maketitle

\begin{abstract}

Numerous aspects and mechanisms of color confinement in QCD are
surveyed. After a gauge--invariant definition of order parameters,
the phenomenon is formulated in the language of field correlators, to
select a particular correlator responsible for confinement. In terms
of effective Lagrangians confinement is viewed upon as a dual
Meissner reffect and a quantitative correspondence is established via
the popular abelian projection method, which is explained in detail.
To determine the field configurations possibly responsible for
confinement, the search is made among the classical solutions, and
selection criterium is introduced.
Finally all facets of confinement are illustrated by a simple example
of string formation for a quark moving in the field of heavy
antiquark.

\end{abstract}

1. Introduction.

2. Definition of confinement and order parameters.

3. Field correlators and confinement.

4. Confinement and superconductivity. Dual Meissner effect.

5. Abelian projection method.

6. Search for classical solutions. Monopoles, multiinstantons and
dyons.

7. Topology  and stochasticity.

8. Conclusions.

9. List of references.
\newpage

 \section{Introduction}

Ten years ago the author has delivered lectures  on confinement at
the  XX LINP Winter School [1]. Different models of confinement were
considered, and the idea was stressed that behind the
phenomenon of confinement is a disorder or stochasticity of vacuum
at large distances.

This phenomenon becomes especially beautiful when  one discovers
that the disorder is due to topologically nontrivial configurations
[2], and the latter could be quantum or classical. In that case one
should look for appropriate classical solutions: instantons,
multiinstantons, dyons etc.

The tedious work of many theorists during last ten years has
given an additional support for the stochastic mechanism, and has
thrown away some models popular in the past, e.g. the dielectric
vacuum model and the $Z_2$ flux model. Those are  reported in [1]
(see also refs.  cited in [1]). At the same  time a new and deeper
understanding of confinement has grown.

It was found that the stochastic mechanism reveals itself on the
phenomenological level as a dual Meissner effect, which was
suggested as a confinement mechanism 20 years ago [3]. During last
years there appeared a lot of lattice data in favour of the dual
Meissner effect and  the so-called method of abelian projection was
suggested [4], which helps to quantify the analogy between the QCD
string and the Abrikosov-Nilsen-Olesen (ANO) string.

At the same time the explicit form of confining configuration --
whether it is some classical solutions or quantum fluctuations --
 is still unclear, some  candidates are being considered
 and active  work is going on.

 The present review is based on numerous data obtained from
 lattice calculations, phenomenology of strong  interaction and
 theoretical studies.

The structure of the rewiew is the following. After formulation of
criteria for confining mechanisms in the next chapter, we
describe in Section 3  the general method of vacuum correlators
(MVC) [5] to characterize confinement as a property of field
correlators.
A detailed comparison of superconductivity and
confinement
in the language of MVC and in the effective Lagrangian formalism of
Ginzburg--Landau type is given in Section 4.

The Abelian Projection (AP) method is introduced in Section 5 and
lattice data are used to establish the similarity of the QCD vacuum
and the dual superconductor medium. At the same time the AP method
reveals the topological properties of confining configurations.

In Section 6 the most probable candidates for such configurations are
looked for among classical solutions. A new principle is introduced
to select solutions which are able to confine when a dilute gas is
formed of those.

An interconnection of topology and stochastic properties of the
QCD vacuum is discussed in Section 7.

In conclusion the search for confining configurations is
recapitulated and  a simple picture of the confining vacuum
and string formation is given as seen from different points of view.
A possible temperature deconfinement scenario is shortly discussed.

\section{ What is confinement?}

 By confinement it is understood the phenomenon of absence in
 physical spectrum those particles (fields) which are present  in
 the fundamental Lagrangian. In the case of QCD it means  that
 quarks and gluons and  in general all colored objects cannot exist
 as separate asymptotic objects.

To be more exact let us note,  that sometimes the massless
particles entering with zero bare masses in the Lagrangian, e.g.
photons or gluons, can effectively acquire mass -- this is the
phenomenon of screening in plasma or in QCD vacuum above the
deconfinement temperature. Due to the definition of confinement
given above, the screened asymptotic color states cannot exist in
the confining phase.

On the other hand in the deconfined phase quarks and gluons can
evolve in the Euclidean (or  Minkowskian) time sepately, not
connected by strings. From this fact it follows that the free
energy for large temperature $t$ is  proportional to $t^4$ as it
usual for the Stefan--Boltzmann gas, and it is natural to call
this phase the quark--gluon plasma (QGP). It turns out, however,
that the dynamics of QGP is  very far from the ideal gas of quarks
and gluons, and  a new characteristic interaction appears there,
called the \underline{magnetic  confinement}  --  this will be
discussed in more detail in conclusions, but now we come back to
the confinement phase.

It is very inportant to discuss confinement in gauge--invariant
terms. Then the physical contents of the confinement phenomenon can
be best understood comparing a gauge--invariant system of an electron
and positron ($+e,-e$) in QED and a system of quark antiquark
($q\bar{q}$) in QCD when no  other charges are present.  When
distance between $+e$ and $-e$ is large the electromagnetic
interaction becomes negligible and the wave function (w.f.) of
$(+e,-e)$ factorizes into a product of individual w.f.  (the same is
true for a gauge--invariant Green's function of ($+e,-e$). Therefore
the  notion of isolated e.m. charge and its individual dynamics makes
sense. In contrast to that in the modern picture of confinement  in
agreement with experiment and lattice data (see below), quark and
antiquark attract each other with the force of appro\-xi\-ma\-te\-ly
14 ton.  Therefore $q$ and $\bar{q}$ cannot separate and individual
dynamics, individual w.f.  and Green's function of a quark
(antiquark) has in principle no sense: quark (or antiquark) is
\underline{confined} by its partner.  This statement can be
generalized to include all nonzero color changes, e.g. two gluons are
confined and never escape each other (later we shall discuss what
happens when additional color charges  come into play).

It is also clear now that the phenomenon of confinement is connected
to the formation of a string between color charges -- the string
gives a constant force at large distances; and the  dynamics of color
charges there without the string is inadequate.

This picture is nicely illustrated by many lattice measurements of
potential between static quarks, Fig. 1.

One can see in Fig. 1  clearly the linear growth of potential
$V(r)=\sigma r$ at large distance $r$; $r\geq 0.25 $ fm.

We shall formulate this as a first property of confinement.

{\it I. The linear interaction between colored objects.}

 To give an exact meaning to this statement, making it possible  to
 check in lattice calculations, it is convenient to introduce the
so--called Wilson loop [6], and to define through it the potential
between color changes. This is done as follows. Take a heavy quark
$Q$ and a heavy    antiquark $\bar Q$ and consider a process,
where the pair $Q\bar Q$ is created at some point $x$, then the
pair separates at distance $r$, and after a period of time $T$ the
pair annihilates at the point $y$. It is clear that trajectories of
$Q$ and $\bar Q$ form a loop $C$ (starting at the point $x$ and
passing through the point $y$ and finishing at the point $x$). The
amplitude (the Green's function) of such a process according to
Quantum Mechanics, is proportional to the phase (Schwinger) factor
$W\sim exp ~ ig \int A_{\mu}j_{\mu}d^4x$, where $A_{\mu}$ is the
total color vector potential of quarks and vacuum fields, and
$j_{\mu}$ -- the current of the pair $Q\bar Q$, depicting its
motion along the loop $C$. Taking into account that $A_{\mu}$ is a
matrix in color space
$(A_{\mu}=A_{\mu}^aT^a,~~a=1,...N^2_c-1,$ and
$T^a=\frac{1}{2}\sigma^a$ for SU(2) and
$T^a=\frac{1}{2}\lambda^a$ for SU(3), where $N_c$ number of colors
$\sigma^a$, $\lambda^a$ -- are Pauli and Gell--Mann matrices
respectively) one should insert an ordering operator $P$ which orders
these matrices along the loop, and the trace operator in color
indices $tr$ , since quarks were created and annihilated in a
white state of the object $Q\bar Q$. Hence one obtains the Wilson
operator for a given field distribution
$W(C)=tr~P~exp~ig~\int_CA_{\mu}dx_{\mu}$ where we have used the
point-like structure of quarks, $j_{\mu}(x)\sim
\delta^{(3)}(x-x(t))\frac{dx_{\mu}}{dt}$

Now one must take into account, that the vacuum fields
$A_{\mu}(x)$ form a stochastic ensemble, and one should average
over it. This is a necessary consequence of the vector character
of $A_{\mu}(x)$ and of the Lorentz--invariance of the vacuum,
otherwise for any fixed function $A_{\mu}(x)$ this invariance
would be violated contrary to the  experiments.

Finally, in field theory in general and dealing  with field vacuum
in particular  it is convenient to use the Euclidean space--time
for the path--integral representation of partition functions,  and
also for  the Monte-Carlo calculations on the lattice. There are
at least two reasons for that: a technical reason; Euclidean path
integrals have a real (and positive-- definite) measure, $
exp(-S_E)$, where  $S_E$ is the Eucledian action, and the
convergence of the path   integrals is better founded (but not
strictly proved in the continuum).

Another, and possibly a deeper reason: Till now all nontrivial
classical solutions in QCD (instantons, dyons etc.) have Eucledian
nature, i.e. in the usual Minkowskian spacetime they describe some
tunneling processes. As for the covalent bonds between atoms due
to the electron tunneling from one atom to another, also the
\underline{Eucledian configurations in gluodynamics and QCD}\\
 \underline{yield
attraction},
i.e. the lowering of the vacuum energy, and therefore are
advantageous for the nonperturbative vacuum reconstruction  (cf
discussion in Conclusion). Therefore everywhere in the review we
shall use the Eucledian space--time, i.e. we go over from
$x_0,A_0$ to the real Euclidean components
$x_4=ix_0,A_4=iiA_0$.

Now we come back to the Wilson loop and relate it to the
potential. To this end we recall that $W$ is an amplitute, or the
Green's function of the $Q\bar Q$ system, and therefore it can be
expressed through the Hamiltonian $H$, namely $W= exp (-HT)$, where
$T$ is the Eucledian time. For heavy quarks the kinetic part of $H$
vanishes and only the $Q\bar Q$ potential $V(r)$ is left, so that
we finally obtain for the averaged (over all vacuum field
configurations denoted by angular brackets) Wilson loop
 \be
 <W(C)>=<tr~ P~
exp~ ig\int_C A_{\mu}dx_{\mu}>= exp(-V(r)T)
 \ee
 where the loop $C$ can be conveniently chosen as a rectangular
 $r\times T$, and confinement corresponds to the linear dependence of
 the potential $V(r)$, $V(r)=\sigma r$, where $\sigma$ is called the
 string tension.

Having said this, one should add necessary elaborations. First, what
happens when other charges are present, e.g. $q\bar{q}$
 pairs are created from the vacuum. In both cases
screening occurs, but whereas for $e\bar{e}$ system nothing crucial
happens at large distances, the $q\bar{q}$ system can be split by an
additional $q_1\bar{q}_1$ pair into two neutral systems of
$q\bar{q}_1$ and $q_1\bar{q}$ which can now separate -- the string
breaks into two pieces.

The same is always true for a pair of gluons which can be screened by
gluon pairs from  the vacuum.

In practice (i.e. in lattice computations) the pair creation
from the vacuum is suppressed even for gluons, [7], one of suppression
factors for quark pairs is $1/N_c$ [8], another one is numerical and
not yet understood, for $N_c=3$ the overall factor is around $0.1$,
as can be seen e.g. in ratio of resonance widths over their masses
$\Gamma/M$.

This circumstance allows one  to see on the lattice the almost
constant force between static $q$ and $\bar{q}$ up to the  distance
of around 1 fm or larger.  This property can be seen in lattice
calculations made with the account of dynamical fermions (i.e.
additional quark pairs), e.g. on  Fig.2 one can see the persistence
of linear confinement in all measured region with accuracy of 10\%.
Therefore we formulate the second property of confinement in QCD,
which should be obeyed by all realistic theoretical models.

{\it II.  Linear confinement between static quarks persists also in
presence of $q\bar{q}$ pairs in the physical region $(0.3 fm \leq r
\leq 1.5 fm$). } Another important comment concerns quark (or gluon)
dynamics at small distances. Perturbative interaction dominates there
because it is singular, which can be seen in the one--gluon--exchange
force of $\frac{\alpha_s}{r^2}$.  Comparing this with the confinement
force mentioned above ($\sigma=0.2 GeV^2\cong 14 ton$), one can see
that perturbative dynamics dominates for $r<0.25 fm$.

At these distances quarks and color charges in general can be
considered  independent, with essentially
perturbative dynamics, which is supported by many successes of
perturbative QCD.

Till now only color charges in fundamental representation of
$SU(N_c)$ (quarks) have been discussed.

 Very surprising results have
been obtained for interaction of static charges in other
representations.  For example adjoint charges, which can easily be
screened by gluons from the vacuum, in lattice calculations are
linearly confined in the physical region $(r\leq 1.5 fm$), Fig. 3.

One can partly understand this property from the  point of view of
large $N_c$: the screening part of potential $V_S\sim \frac{1}{r} exp
(-\mu r) $ is suppressed by the factor $1/N_c^2$ as compared with
linear part, but wins at large distances [7]
\be
<W_{adj}(C)>=C_1exp(-\sigma_{adj} r T)+\frac{C_2}{N_c^2} exp
(-V_s(r)T)
\ee

The same property holds for other charge representations [10], and
moreover the string tension $\sigma(j)$ for a given representation $j$
satisfies an approximate relation [7]
\be
\frac{\sigma(j)}{\sigma{(fund)}}=\frac{C_2(j)}{C_2(fund)}
\ee
where  $C_2(j)$ is the quadratic Casimir operator,
$$C_2(adj,N_c)=N_c,~~
C_2(fund,N_c)=\frac{N_c^2-1}{2N_c}$$

Correspondingly we formulate the third property of confinement.

{\it III. Adjoint and other charges are effectively confined in the
physical  region $(r\leq 1.5 fm$) with string tension satisfying
(2).}

We conclude this section with two remarks, one concerning
interrelation of confinement and  gauge invariance,  another about
many erroneous attempts to define confinement through some specific
form of nonperturbative gluon and quark propagator.

Firstly, gauge invariance is absolutely necessary to study properly
the mechanism of confinement. It requires that any
gauge--noninvariant quantity, like quark or gluon propagator,
3--gluon vertex etc. vanish when  averaged over all gauge copies. One
can fix the gauge only for the gauge--invariant amplitude, otherwise
one lacks important part of dynamics,  that of confinement.

As one popular example one may consider the "nonperturbative" quark
propagator, which is suggested in a form without poles at real
masses. As was told above, this propagator has no sense, since on
formal level it is gauge--noninvariant and vanish upon averaging,
and on  physical grounds, propagator of a colored object cannot be
considered separately from other  colored partner(s), since it is
connected by a string to it, and this string (confinement) dynamics
dominates at large distances. The same can be told about the
"nonperturbative" gluon propagator  behaving like $1/q^4$ at small
$q$, which in addition imposes wrong singular nonanalytic behaviour
of two--gluon--glueball Green's function.

Likewise the well--known Dyson--Schwinger  equations (DSE) for
one--particle Green's functions cannot be used for QCD in the
confined phase, because again they are i) gauge--noninvariant ii)
there should be a string  connected to any of propagators of DSE,
omitted there.

More subtle formally, but basically the same is situation with
Bethe--Salpeter equation (BSE) on the fundamental level, i.e, when
the kernel of BSE contains colored gluon or quark exchanges: any
finite approximation (e.g. ladder--type) of the kernel violates gauge
invariance and looses confinement. When averaging over all vacuum
fields is made in the amplitude on the other hand, the resulting
effective interaction can be treated approximately in the framework
of BSE.
Their value if any might lie in phenomenological applications and not
in fundamental understanding of the confinement, which is the primary
purpose of the present review.

Let us turn now to the order parameters defining the confining phase.
To distinguish between confined and deconfined phase several order
parameters are used in absence of dynamical (sea) quarks. One is the
introduced above Wilson loop (1). Confinement is defined as the phase
where the area law (linear potential) is valid for large contours,
whereas in the  deconfining phase the perimeter law appears.

Another, and sometimes practically more convenient for nonzero
temperature $t$, is the order parameter called the Polyakov line:
\be
<L(\vec{x})>=\frac{1}{n}<tr~ P~ exp~ ig\int^{\beta}_0A_4dx_4>,
~~\beta=1/t
\ee
Since $<L>$ is connected to the free energy $F$ of isolated
color charge, $<L>=exp(-F/T)$, vanishing of $<L>$ at
$T<T_C$ means that $F$  is infinite in the confined phase. On the
other hand, vanishing of $<L>$ is connected to the $Z(N_c)$
symmetry, which is respected in the confining vacuum and broken in
the deconfined phase. As for the Wilson loops, Polyakov lines can
be defined both for fundamental and adjoint charges; the first in
absence of dynamical quarks vanish rigorously in confined phase,
while adjoint lines are very small  there.
When dynamical quarks are admitted in the vacuum, both Wilson loop
and Polyakov line are not order parameters, strictly speaking, there
is no area law for large enough Wilson loops and $<L>$ does not
vanish in the confined phase. However, for large $N_c$, and
practically even for $N_c=2,3$ these quantities can be considered as
approximate and useful order parameters even in presence of dynamical
quarks, as will be seen in next sections.

From dynamical point of view the (approximate)  validity of area law
(linear potential) for any color charges even in presence of
dynamical quarks at distances $r\leq 1.5 fm$ means that strings are
formed at these distances and string dynamics defines the behaviour
of color charges in the most physically interesting region.

  \section{Field correlators and confinement picture}

As can be seen in the area law (1) the phenomenon of confinement
necessarily implies appearance of a new mass parameter in the theory
-- the string tension $\sigma$ has dimension [mass]$^2$.

Since perturbative QCD depends on the mass scale only due to the
renormalization through $\Lambda_{QCD}$, and using to the
asymptotic freedom (and renormalization group properties) one can
express the coupling constant $g(\Lambda)$ at the scale $\Lambda$
through $\Lambda_{QCD}$ as $g^2(\Lambda)= \frac{(4\pi)^2}{\beta_0
ln\frac{\Lambda^2}{\Lambda^2_{QCD}}},~~\beta_0=\frac{11}{3}
N_C-\frac{2}{3} n_f,$
 one can write
 \be
  \sigma= const
\Lambda^2_{QCD} \sim \Lambda^2 exp
(-\frac{16\pi^2}{\beta_0g^2(\Lambda)})
 \ee
 where $\Lambda$ is the cut--off momentum.
 It is clear from (5) that $\sigma$ cannot be
obtained from the perturbation series, hence the source of $\sigma$
and of the whole confinement phenomenon is purely nonperturbative.
Correspondingly one should admit in the QCD vacuum the
nonperturbative component and split the total gluonic vector
potential  $A_{\mu}$ as \be A_{\mu}=B_{\mu}+a_{\mu} \ee where
$B_{\mu}$ is nonperturbative and $a_{\mu}$ -- perturbative part.  As
for $B_{\mu}$, it can be\\
a) quasiclassical, i.e. consisting of a superposition of classical
solutions like instantons, multiinstantons, dyons etc. This
possibility will be discussed below in  Section 6.  \\
b) purely
quantum (but nonperturbative). The picture of Gaussian stochastic
vacuum gives an example, which is discussed below.

It is important to stress at this point, that the formalism of field
correlators, given below in this chapter, is of general character and
allows to  discuss both situations a) and b), quasiclassical and
stochastic.  In the first case, however, some modifications are
necessary which will be introduced at the end of the chapter.
As it was mentioned above, confinement implies   the string formation
between  color charges. To understand  how string is related to
field correlators, consider a simple example of a nonrelativistic
quark moving at a distance $r$ from  a heavy antiquark fixed  at
the origin. As it is known from quantum mechanics, the quark
Green's function is proportional to the phase integral along its
trajectory $C$
$$
G(\vec r, t) \sim <exp~ig \int_C A_{\mu}(\vec r(t'),t') d z_{\mu}>
$$
where the averaging is over all vacuum configurations. It is
convenient to express
 $A_{\mu}$ though the field strength
$F_{\mu\nu}$, since the latter would be the basic stochatic
quantities, and this can be done e.g. using the Fock--Schwinger gauge
 $$
A_{\mu}(x)= \int^x_0 F_{\nu\mu}(u) \alpha(u) du_{\nu},~~
\alpha(u)=\frac{u}{x} $$
Hence in the lowest order one obtains
$$ G(\bar r,t)\sim 1 - \frac{g^2}{2} \int
d\sigma_{\nu\mu}(u)
  d\sigma_{\nu'\mu'}(u')<F_{\nu\mu}(u)F_{\nu'\mu'}(u')>+...  $$
  where notations are used such that
  $d\sigma_{\nu\mu}=\alpha(u) dx_{\mu} du_{\nu}$ .

 On the other hand one inroduce the potential
  $V(r)$ acting on the quark
   $$ G\sim exp (-\int
 V(r, t') dt') \sim 1 - \int V (r,t')dt' $$
 the string formation implies that
  $V(r,t) $  is proportional to $r$ and this depends, as we see, on
  the field correlators $<FF>$.

               For the exact Lorentz--invariant treatment let us introduce
 gauge--invariant field correlators (FC) and express the Wilson loop
  average through FC. This is done using the nonabelian Stokes
theorem [11]
 \be <W(C)>=<\frac{1}{N_C} P tr ~exp~ig~ \int_c A_{\mu}
dx_{\mu}>= \ee $$\frac{1}{N_C}<P ~tr~exp~ig~\int_S
d\sigma_{\mu\nu}F_{\mu\nu}(u,z_0)> $$ where we have defined \be
F_{\mu\nu}(u,z_0) = \Phi(z_0,u) F_{\mu\nu} (u) \Phi(u,z_0), \Phi(x,y)
= P \exp ig \int^x_y A_{\mu} dz_{\mu}
\ee
 and integration in (7) is
over the surface $S$ inside the contour $C$, while $z_0$ is an
arbitrary point, on which $<W(C)>$ evidently does not depend.  In the
Abelian case the parallel transporters $\Phi(z_0,u)$  and
 $\Phi(u,z_0)$ cancel and one obtains the usual Stokes theorem.

Note that the nonabelian Stokes theorem, eq. (7), is gauge invariant
even before averaging over all vacuum configurations -- the latter is
implied by the angular brackets in (7).

One can now use the cluster expansion theorem [12] to express the
r.h.s. of (7) in terms of $FC$, namely [5]
\be
<W(C)> = \frac{tr}{N_C} \exp \sum^{\infty}_{n=1} \frac{(ig)^n}{n'}
\int d\sigma(1)d\sigma(2)... d\sigma(n)\ll F(1)... F(n)\gg
\ee
where lower indices of $d\sigma_{\mu\nu}$ and $F_{\mu\nu}$ are
suppressed and $F(k) \equiv F_{\mu_k \nu_k}(u^{(k)}, z_0)$.

Note an important simplification -- the averages $\ll F(1) ...
F(n)\gg$ in the color symmetric vacuum are proportional to the unit
matrix in color space, and the ordinary operator $P$ is not needed
any more.

Eq. (9) expresses Wilson loop in terms of gauge invariant $FC$,
also called cumulants [12], defined in terms of $FC$ as follows:
\be
\ll F(1) F(2) \gg = < F(1)F(2)> - <F(1)><F(2)>
\ee
$$
\ll F(1) F(2) F(3) \gg = <F(1)F(2)F(3)>-\ll F(1)F(2)\gg
<F(3)>-
$$
$$
-<F(1)>\ll F(2)F(3)\gg - <F(2)>\ll F(1)F(3)\gg-<F(1)><F(2)><F(3)>
$$
Let us have a look at the lowest cumulant
\be
 <F(x)
\Phi(x,z_0) \Phi(z_0,y) F(y) \Phi(y,z_0) \Phi(z_0,x)> \ee
It depends
not only on $x,y$ but also on the arbitrary point $z_0$.  In case
when a classical solution (a dyon or instanton) is present it is
convenient to place $z_0$ at its center, and then $z_0$ acquires a
clear physical meaning. We shall investigate this case in detail in
chapter 6, but now we consider the limit of stochastic vacuum,
when the expansion (9) is particularly useful. To this end
consider parameters on which a generic cumulant $\ll F(1) ...
F(n)\gg$ depends. When all coordinates $u^{k}$ coincide with $z_0$,
one obtains condensate $\ll (F_{\mu_n\nu_n}(0))^n \gg$, to which we
assign an order of magnitude $F^n$. The coordinate dependence can
be characterized by the gluon correlation length $T_g$, which is
assumed to be of the same order of magnitude for all cumulants. Then
the series in (9) has the following estimate
 \be <W(C)> =
\frac{tr}{N_c} \exp \sum^{\infty}_{n=1} \frac{(ig)^n}{n!} F^n
T_g^{2(n-1)} S \ee
 where $S$ is the area of the surface inside  the
contour $C$. To obtain the result (12) we have taken into account
that in each cumulant
\be
 \ll F(x^{(1)}, z_0) F(x^{(2)}, z_0) ... F(x^{(n)},z_0)
\gg \ee whenever $x$ and $y$ are close to each other.
 $$ \mid x-y
\mid \ll \mid x - z_0 \mid,\; \mid y - z_0 \mid\;,$$ the dependence
on $z_0$ drops out, therefore in (13) in a generic situation where
all distances $\mid x^{(i)} - x^{(j)} \mid \sim T_g \ll \mid x^{(i)}
- z_0 \mid\;,\; \mid x^{(j)} - z_0 \mid$ one can omit dependence on
$z_0$.

The expansion in (12) is in powers of $(FT^2_g)$, and when this
parameter is small,
\be
FT^2_g \ll 1
\ee
one gets the limit of Gaussian stochastic ensemble where the lowest
(quadratic in $F_{\mu\nu}$) cumulant is dominant.

In the same approximation (e.g. $T_g \to 0$ while $<F^2_{\mu\nu}>$ is
kept fixed) one can neglect in this cumulant the $z_0$ dependence,
using the equivalent (effective) form  of (11)
\be
D_{\mu\nu\lambda\sigma} \equiv \frac {1}{N_C}
tr <F_{\mu\nu}(x) \Phi(x,y)F_{\lambda\sigma}(y)\Phi(y,x)>
\ee
The form (15) has a general decomposition in terms
of two Lorentz scalar functions $D(x-y)$ and $D_1(x-y)$ [5]
 \begin{eqnarray}
D_{\mu\nu\lambda\sigma} &=&
(\delta_{\mu\lambda}\delta_{\nu\lambda} -
\delta_{\mu\sigma}\delta_{\nu\lambda}){\cal{D}}(x-y) + \\
\nonumber
&+& \frac{1}{2} \partial_{\mu} \{[(h_{\lambda} \cdot
\sigma_{\nu\delta} - h_{\delta}\sigma_{\nu\lambda}) + ... ]
{\cal{D}}_1(x-y)\}
\end{eqnarray}
 Here the ellipsis implies terms obtained by permutation of indices.
 It is important that the second
term on the r.h.s. of (16) is a full derivative by construction.

Insertion of (16) into (9)  yields the area law of
Wilson loop with the string tension $\sigma$ $$<W(C)>= exp (-\sigma
S_{min})$$
 \be
  \sigma=\int D(x)d^2x(1+O(FT_g^2))
  \ee
  where $O(FT_g^2)$
stands for the contribution of higher cumulants, and $S_{min}$ is the
minimal area for contour $C$.

Note that $D_1$ does not enter $\sigma$, but gives rise to the
perimeter  term and higher order curvature terms. On the other
hand  the lowest order perturbative QCD contributes to $D_1$ and not
to $D$, namely the one--gluon--exchange  contribution is
\be
D_1^{pert}(x)=\frac{16 \alpha_s}{3\pi x^4}
\ee

 Nonperturbative parts of $D(x)$ and $D_1(x)$ have been computed on the lattice [13]
 using the looking method, which suppresses perturbative
 fluctuations,
 and are shown in Fig.4. As one can see in Fig.4, both functions are well
 described by an  exponent in the measured region, and $D_1(x)\sim \frac{1}{3}
 D(x)\sim exp (-x/T_g)$ , where
 $T_g\sim 0.2 fm$.
 The smallness of $T_g$ as compared to hadron size confirms the approximations
 made before, in particular the stochasticity condition
 (14). One should also take into account that $F$ in (14) is an effective
 field in cumulants, which vanish when vacuum insertion
 is made and therefore can be small as compared with $F$ from
 the gluonic condensate.

The representation (16) is valid both for abelian and nonabelian
theories, and it is interesting whether the area law and nonzero
string tension obtained in (17) could be valid also for QED (or
$U(1)$ in lattice version). To check it let us apply the operator
$\frac{1}{2}\varepsilon_{\mu\nu \alpha\beta}\frac{\partial}{\partial
x_{\alpha}}$ to both sides of (15-16) [].

In the abelian case, when $\Phi$ cancel in (15),  one   obtains
(the term with $D_1$ drops out)
\be
\partial_{\alpha}<\tilde{F}_{\alpha\beta}(x)F_{\lambda\sigma}{(y)}>=
\varepsilon_{\lambda\sigma\gamma\beta}\partial_{\gamma}D(x-y)
\ee
where
$\tilde
F_{\alpha\beta}=\frac{1}{2}\varepsilon_{\alpha\beta\mu\nu}F_{\mu\nu}.$

If magnetic monopoles
 are present in Abelian theory (e.g.  Dirac monopoles) with the
 current $\tilde{j}_{\mu}$, one has \be
\partial_{\alpha}\tilde{F}_{\alpha\beta}(x)=\tilde j_{\beta}(x) \ee
In absence of magnetic monopoles (for pure QED) the abelian Bianchi
identity (the second pair of Maxwell equation) requires that
\be
\partial_{\alpha}\tilde{F}_{\alpha\beta}\equiv  0
\ee
Thus for QED (without magnetic monopoles) the function $D(x)$
vanishes due to (19) and hence confinement is absent, as observed
in nature.

In the lattice version of $U(1)$ magnetic monopoles are present (as
lattice artefacts) and the lattice formulation of our method would
predict the confinement regime with nonzero string tension, as it is
observed in Monte--Carlo calculations [14].

The latter can be connected through $D(x)$ to the correlator of
magnetic monopole currents. Indeed multiplying both sides of (19) with
$\frac{1}{2}
\varepsilon_{\lambda\sigma\gamma\delta}\frac{\partial}{\partial y_{
\gamma}}$ one obtains
\be
<\tilde{j}_{\beta} (x) \tilde{j}_{\delta} (y)> =
(\frac{\partial}{\partial x_{\alpha}} \cdot \frac{\partial}{\partial
y_{\alpha}} \cdot \delta_{\beta\delta} - \frac{\partial}{\partial
x_{\delta}} \frac{\partial}{\partial y_{\beta}})  D(x-y)
\ee
The form of Eq.(22) identically satisfies monopole current
conservation:
applying $\frac{\partial}{\partial x_{\beta}}$ or
$\frac{\partial}{\partial y_{\delta}}$ on both sides of (22) gives
zero.

It is interesting to note, that confinement (nonzero $\sigma$ and
$D$, see  Eq. (17)) in $U(1)$ theory with monopole currents
occurs not due to average monopole density $<\tilde{j}_4(x)>$, but
rather due to a  more subtle feature -- the  correlator of monopole
currents (22), which can be nonzero for the configuration where
$<\tilde{j}_4(x)> = 0$.  The latter is fulfilled for the system with
equal number of monopoles and antimonopoles.

Note that our analysis here is strictly speaking applicable when
stochasticity condition (14) is fulfilled. Therefore the case of
monopoles with Dirac quantization condition is out of the region of
(14) and needs some  elaboration to be discussed later in this
chapter.

Let us turn now to the nonabelian case, again assuming stochasticity
condition (14), so that one can keep only the lowest cumulant
(15-16).

Applying as in the Abelian case the operator $\frac{1}{2}
\varepsilon_{\mu\nu\alpha\beta} \frac{\partial}{\partial x_{\alpha}}$
to the r.h.s. of (15-16), one obtains [15]
\begin{eqnarray}
<D_{\alpha} \tilde{F}_{\alpha\beta} (x) \Phi(x,y) F_{\lambda\sigma}
(y) \Phi(y,x)> + \Delta_{\beta\lambda\sigma} (x,y) &=& \\
\nonumber
&=& \varepsilon_{\alpha\beta\lambda\sigma} \frac{\partial}{\partial
x_{\alpha}} D(x-y)
\end{eqnarray}

The term with $D_1$ in (23) drops out as in the Abelian case; the
first term on the l.h.s. of (23) now contains the nonabelian
Bianchi identity term which should vanish also in presence of dyons
(magnetic monopoles) --  classical solutions of Yang-Mills theory,
i.e. one has \be D_{\alpha} \tilde F_{\alpha\beta}(x) = 0. \ee

It is another question, whether or not in lattice formulation one can
violate (24) in the definition of lattice artefact monopoles,
similarly to the  Abelian case. We shall discuss this topic when
studying lattice results on  Abelian projected monopoles in Section
5.
To conclude discussion of (23) one should define
$\Delta_{\beta\delta\sigma}$; the latter appears only in the
nonabelian case due to the shift of the straight-line contour $(x,y)$
of the correlator (15) into the position $(x+ \delta x, y)$, which is
implied by differentiation
 $\frac{\partial}{\partial x_{\alpha}}$. This
"contour differentiation" is well known in  literature [16] and
leads to the answer
\be
\Delta_{\beta\lambda\sigma}(x,y)=ig \int^x_y
dz_{\rho}\alpha(z)<tr[\tilde{F}_{\alpha\beta}(x)\phi(x,z)
F_{\alpha \rho}(z)\phi(z,y)F_{\lambda\sigma}(y) \phi(y,x)-
 \ee
$$-\tilde{F}_{\alpha\beta}(x) \phi(x,y) F_{\lambda\sigma}(y)
 \phi(y,z)F_{\alpha\rho}(z)\phi(z,x)]>.
$$
 Especially simple form of
(23) occurs when using (23) and tending $x$ to $y$; one obtains
[15] \be \frac{dD(z)}{dz^2}|_{z=0}=
\frac{g}{8}f^{abc}<F^a_{\alpha\beta}(0)
F^b_{\beta\gamma}(0)F^c_{\gamma\alpha}(0)>.
\ee
Thus confinement (nonzero $\sigma$ due to nonzero $D$)
occurs in nonabelian case due to the purely nonabelian correlator\\
 $<tr
F_{\alpha\beta}F_{\beta\gamma}
F_{\gamma\alpha}>=3<trE_iE_jB_k>\varepsilon_{ijk}$.

To see the physical
 meaning of this correlator, one can  visualize
magnetic and electric field strength lines (FSL) in the space. Each
magnetic monopole is a source of FSL, whether it is a real object
(classical solution or external object like Dirac monopole) or
lattice artefact.

In nonabelian theory these lines may form branches, and e.g. electric
FSL may emit a magnetic FSL at some point, playing the role of
magnetic monopole at this  point. This is what exactly nonzero
triple correlator\\ $<TrFFF>$ implies. Note that this could be a
purely quantum effect, and no real magnetic monopoles are necessary
for this mechanism of confinement.

Till now we have discussed confinement in terms of the lowest
cumulant -- $D(x)$, which is justified when stochasticity condition
(14) is fulfilled and $D(x)$ gives a  dominant contribution. Let
us now turn on other  terms in the cluster expansion (9). It is
clear that the general structure of higher cumulants is much more
complicated than (16), but there is always present a
Kronecker-type term $D(x_1, x_2, ..., x_n)\prod \delta_{\mu_l \mu_k}$
similar to $D(x-y)$ in (16) and other terms containing
derivatives and coordinate differencies like $D_1$. The term $D(x_1,
x_2, ... , x_n)$ contributes to string tension, and application of the
same operator $\frac{1}{2} \varepsilon_{\mu\nu\alpha\beta}
\frac{\partial}{\partial x_{\alpha}}$ again reveals the nonabelian
Bianchi identity term (24) and an analog of
$\Delta_{\beta\lambda\sigma}$ in (23). This means that the string
tension in the  general case is a sum
\be
 \sigma = \sum^{\infty}_{n=2}
\sigma^{(n)}\;, \; \sigma^{(n)} =\frac{ g^n}{n!}
\int\ll F(1) F(2) ... F(n)\gg
d\sigma (2) ... d\sigma(n)
\ee
 When the stochasticity condition
(14) holds, the lowest term, $\sigma^{(2)}$, dominates in the
sum; in the general case all terms in the sum (27) are important.
The most important case of such a situation is the discussed above
the case of quasiclassical vacuum, which we now shortly discuss,
shifting the detailed discussion to the Section 6.

For a dilute gas of classical solutions the role of vacuum
correlation length $T_g$ is played by the size of the solution
$\rho$. For the correlator $<F(x_1) ... F(x_n)>$ the essential
nonzero result occurs when all $x_1, ... x_n$ are inside the radius
$\rho$
of the solution (e.g. dyon or instanton).

On the other side the typical $F$  of the solution,
e.g.: $F = F_{\mu\nu}(0)$, is connected to $\rho$ by the value of
topological charge, e.g. for instanton
\be
Q = \frac{g^2}{32\pi^2} \int d^4x F^2_{\mu\nu}
\ee
Since $Q$ is an integer one immediately obtains that
\be
(g F \cdot \rho^2)^2 \sim n\;, \;\; n = 1,2 ...
\ee

The same estimate holds for the dyon, solution which is discussed in
Section 6.

Hence the series (12) and (27) has a parameter of expansion
$(gFT^2_g)$ of order of unity and may not converge. A more detailed
analysis of the gas of instantons and magnetic monopoles shows that
the string tension series (27) for instantons (for simplicity
centers of instantons and monopoles were taken on the plane (12)
of the Wilson loop) looks like [17]
 \be \sigma=\rho_0(1-<cos \beta>);
 ~~\beta=g\int F_{12}(z)d^2z=2\pi,~~ \sigma=0
 \ee
  while for
magnetic monopoles
\be
\beta=\pi,~~~ \sigma =2\rho_0
,
 \ee
and $\rho_0$ is the surface density of instantons (monopoles).

Note that $\sigma^{(2)}= \rho_0\frac{<\beta^2>}{2}$ in both cases is
positive, and for instantons the total sum for $\sigma$ vanishes
while for monopoles $\sigma$ is nonzero. As we shall see below this
is in agreement with the statement in chapter 6: all topological
charges (29) yield flux through Wilson  loop equal to $2\pi Q$, and
for (multi)instantons with $Q = n = 1,2,...$ Wilson loop is $W = \exp
2\pi Q i = 1$, and no confinement results for the dilute gas of such
solutions. For magnetic monopoles (dyons) topology is different and
elementary flux is equal to $\pi$, bringing confinement for the
dilute gas of monopoles in agreement with (31).

Thus the lowest cumulant $<F(x) \Phi F(y)\Phi>$ might give a
misleading result in case of quasiclassical vacuum, and one should
sum up all the series to get the correct answer as in
(30-31).  Therefore to treat the vacuum containing
topological charges one should separate the latter and write their
contribution explicitly, while the rest -- quantum fluctuations with
$F T^2_g \ll 1$ -- can be considered via the lowest cumulants. An
example of such vacuum with instanton gas and  confining
configurations was studied in [18] to obtain chiral symmetry  breaking;
this work demonstrates the usefulness of such an approach.

We conclude this chapter with discussion of confinement for charges in
higher representations. As it was stated in the previous chapter, our
definition of confinement based on lattice data, requires the linear
potential between static charges in any representation, with string
tension proportional to the quadratic Casimir operator.

Consider therefore the Wilson loop (1) for the charge in some
representation; the latter was not specified above in all eqs.
leading  to (27). One can write in general
\be
A_{\mu}(x)=A_{\mu}^aT^a, tr(T^aT^b)=\frac{1}{2}
\delta^{ab}
\ee
Similarly to (9) one has for the representation $j=(m_1,m_2...)$ of
the group $SU(N)$ with dimension $N(j)$
 \be <W(C)=\frac{1}{N(j)}tr_j
exp \sum^{\infty}_{n=1} \frac{(ig)^n}{n!} \int
d\sigma(1)...d\sigma(n)\ll F(1)...F(n)\gg \ee and by the usual
  arguments one has Eq.(27) .

              Due to the color neutrality of the vacuum each cumulant
              is proportional to the unit matrix in the color space,
e.g. for the lowest cumulant one has
\be
<F(1)F(2)>_{ab}=<F^c(1)F^d(2)>T^c_{an}T^d_{nb}=
\ee
$$
=<F^e(1)F^e(2)>\frac{1}{N^2_c-1}
T^c_{an}T^c_{nb}=\Lambda^{(2)} C_2(j)\cdot
\hat{1}_{ab},
$$
where we have used the definition
\be
T^cT^c=C_2(j)\hat{1}
\ee
and introduced a constant not depending on representation,
\be
\Lambda^{(2)}\equiv \frac{1}{N^2_C-1}<F^e(1)F^e(2)>,
\ee
and also used the color neutrality of the vacuum,
\be
<F^c(1)F^d(2)>=\delta_{cd}\frac{<F^e(1)F^e(2)>}{N^2_C-1}
\ee
For the next -- quartic cumulant one has
\be
\ll F(1)F(2)F(3)F(4)\gg_{\alpha\varepsilon}=
\ll F^{a_1}(1)F^{a_2}(2)F^{a_3}(3)F^{a_4}(4)\gg \times
\ee
$$
\times
T^{a_1}_{\alpha\beta}T^{a_2}_{\beta\gamma}T^{a_3}_{\gamma\delta}
T^{a_4}_{\delta\varepsilon}=
\Lambda^{(4)}_1(C_2(j))^2\delta_{\alpha\varepsilon}+\Lambda_2^{(4)}
(T^{a_1}T^{a_2} T^{a_1}T^{a_2})_{\alpha\varepsilon}
$$
Thus one  can see in the quartic cumulant a higher order of quadratic
Casimir and higher Casimir operators.

The string tension for the representation $j$ is the coefficient of
the diagonal element in (34) and (38)
\be
\sigma(j)=C_2(j)\int\frac{g^2\Lambda^{(2)}}{42}d^2x+O(C^2_2(j))
\ee
where the term $O(C^2_2(j))$ contains higher degrees of $C_2(j)$ and
higher Casimir operators.

Comparing our result (39) with lattice data [10] and Fig.3 one can
see that the first quadratic cumulant should be dominant as it
ensures proportionality of $\sigma(j)$ to the quadratic Casimir
operator.

 Another interesting and important check of the dominance of
 bilocal correlator (of the Gaussian stochasticity) is the
  calculation of the QCD
 string profile, done in [19]. Here the string profile means the
 distribution $\rho_{11}$ of the
 longitudinal component of colorelectric field as a function
 of distance $x_{\bot}$ to the string axis. This distribution can
 be expressed through integral of functions $D(x)$, $D_1(x)$ [18], and
  with measured values of $D$, $D_1$ from [13] one can  compute
  $\rho_{11}(x_{\bot})$ and compare it with independent measurements.
  This comparison was done in [19] and shown in Fig. 5.  One can see
   a good agreement of the computed $\rho_{11}(x_{\bot})$ (broken
   line) with "experimental " values. Thus MVC gives a good
   description of data in the simplest (bilocal) approximation, even
   for such delicate characteristics as field
   distributions inside the string.

\section{Dual Meissner mechanism, confinement and superconductivity}

The physical essence of the confinement phenomenon is the formation
of the string between the  probing charges introduced into the
vacuum, which in turn means that the electric field distribution is
drastically changes from the usual dipole picture (for the empty
vacuum) and is focused instead into a string picture in the confining
vacuum. From the point of view of macroscopic electrodynamics of
media, this effect can be described introducing dielectric function
$\varepsilon (x)$, and electric induction $\vec{D}(x)$ together
with  electric field $\vec{E}(x)$,
$\vec{D}(x)=\varepsilon(x)\vec{E}(x)$.

One can then adjust $\varepsilon (x)$, or better $\varepsilon(D)$ or
$\varepsilon(E)$ to obtain the string formation. Another possibility
is to choose the effective action as a function of $F^2_{\mu\nu}$ in
such  a way, as to reproduce string--type distribution.

This direction was reviewed in [1], the main conclusion  reached long
ago [20] was that the physical vacuum  of QCD considered as a medium
could be called a pure diaelectric, i.e. $\varepsilon=0$  far from
probing charges [21].

It is also shown, that it is possible to choose $\varepsilon (E)$ in
such a way as to obtain a string of constant radius [21] or with radius
slowly dependent on length . We shall not follow these results
below, as well as results of the so--called dielectric model [1,21],
referring the reader to the mentioned literature.

Instead we focus in this and following chapter on another approach,
which has proved to be fruitful during last years -- confinement as
dual Meissner effect [3] and ideologically connected to it  abelian
projection method [4].

The physical idea used by 'tHooft  and Mandelstam [3] is the analogy
between the Abrikosov  string formation in type II-- superconductor between
magnetic  poles and  proposed string formation between color electric
charges in QCD.

We shall study this   analogy here from several different point of
view:

1) energetics of the vacuum -- minimal free energy of the vacuum

2) classical equations of  motion --Maxwell and London equations

3) condensate formation and symmetry breaking

4) vacuum correlation functions of fields and currents.

We consider in this chapter the 4d generalization of the
Ginzburg-Landau model of superconductivity, which is called the
Abelian Higgs model with the Lagrangian [22]
\be
{\cal{L}}=-\frac{1}{4}F^2_{\mu\nu} -| D_{\mu}\varphi|^2-\frac{\lambda}{4}
(|\varphi|^2-\varphi^2_0),~~
D_{\mu}=\partial_{\mu}-ieA_{\mu}.
\ee

This model is known to possess  classical
solutions--Nielsen--Olesen strings , which are 4d generalization of
the Abrikosov strings, occurring in type II superconductor. The latter
can  be  described by the Ginzburg--Landau Lagrangian, when coupling
constants are chosen correspondingly [23].

The Lagrangian (40) combines two fields: the electromagnetic  field
($A_{\mu},F_{\mu\nu})$ -- which will be analogue of gluonic  field of
QCD, and the complex Higgs field $\varphi (x)$, which describes the
amplitude of the Cooper--paired  electrons in a superconductor. When
$\lambda\to\infty$ the wave functional $\Psi\{A_{\mu},\varphi(x)\}$
has a strong maximum around $\varphi(x)=\varphi_0$, which means
formation of the condensate of Cooper  pairs of amplitude
$\varphi_0$.

Let us look more closely at the model (40) discussing points 1)--4)
successively.

1) The energy density corresponding to the (40) is
\be
\varepsilon=\frac{\vec{E}^2+\vec{B}^2}{2}+|\vec{D}\varphi|^2+
|D_0\varphi|^2+ \frac{\lambda}{4}
(|\varphi|^2-\varphi^2_0)^2
\ee
From (41) it is clear  that in absence of external sources and for
large $\lambda$ the lowest (vacuum) state corresponds to
$\varphi=\varphi_0=const$, and $\vec{E}=\vec{B}=0$. For future
comparison with QCD it is worthwhile to stress that formation of the
condensate $<F^2_{\mu\nu}>$ is not advantageous for large $\lambda$,
since the mixed term $|A_{\mu}\varphi|^2$ will give very large
positive contribution: \underline{condensate of electric}
\underline{ charges
$\varphi$ repels and suppresses everywhere electromagnetic field
$F_{\mu\nu}$}.
The same situation occurs in the nonabelian Higgs --  the Georgi --
Glashow model: there appear two phases depending on values of
$\lambda , g$ and for large $\lambda$ the deconfined phase with
$\varphi=\varphi_0$ persists. Here the word deconfined means that
external (color) electric charges are not confined, while magnetic
monopoles can be confined.
     For smaller values of $\lambda$ one can reach
     a region where it is advantageous
      to form the condensate $<F^2_{\mu\nu}>$
       and then possibly also nonzero $D(x)$.
       As we discussed in Section 3 then there is a possible
       confinement of colorelectric charges in this phase.  But let
        us come back to the Abelian Higgs model and the limit of
        large $\lambda$, which is of primary interest to us.

In this  case  it is advantageous to form the
condensate of electric charges  $(\varphi(x)=\varphi_0$),
condensate of electromagnetic field is suppressed, and one
obtains Abrikosov-- Nielsen--Olesen (ANO) strings connecting
magnetic poles -- this is the confinement of magnetic monopoles and
mass generation of  e.m. field -- leading to the deconfinement
of electric charges.

In the dual picture one would assume that condensate of magnetic
charges (monopoles) would help to create strings of electric field,
connecting (color) electric charges, yielding confinement of the
latter. Our tentative guess is that the condensate of magnetic
monopoles (dyons) in gluodynamics is associated with the gluonic
condensate $<F^2_{\mu\nu}>$.

2) We now discuss the structure of the ANO strings from the point of
view of classical equations of motion. The Maxwell equations are
\be
rot\vec{B}=\vec{j}
\ee
where $\vec{j}$ is the microscopic current of electric charges,
including the condensate of Cooper pairs.

To obtain a closed equation for $\vec B$ one needs the specific
feature of superconductor in the form of Londons equation
\be
rot \vec {j} = \delta^{-2} \vec B
\ee

The latter can be derived from the Ginzburg--Landau--type Lagrangian
(40).
  Indeed writing the current for the Lagrangian (40) in the usual way, we have
\be
j_{\mu}=ie(\phi^+\partial_{\mu}\phi-\partial_{\mu}\phi^+\phi)-2
e^2A_{\mu}|\phi|^2
\ee
Assume that there exists a domain where
 $\phi$  is already constant, and
 $A_{\mu}$ is still nonzero (we shall define this region later in a better way),
 and apply the rot operation to both sides of (44). Then one obtains the London equations
 (43) and
 $\delta$  is defined by
 \be
 \delta^{-2}=2e^2\phi^2_0
 \ee

Insertion of (43) into (42) yields equation for $\vec{B}$
\be
\Delta\vec{B}-\delta^{-2}\vec{B}=0
\ee
Solving (46) one obtains  the exponential fall--off of $\vec{B}$ away
from the center of the ANO string,
 \be
B(r)=const K_0(r/\delta),~~B(r)\sim
exp(-r/\delta),~~r\gg\delta \ee
From (47) it is clear that
 $\delta^{-1}$   is thephoton  mass, generated by the Higgs
 mechanism.  On the other hand the field $\phi$ has its own
 correlation length $\xi$,  connected to the mass of quanta of the
  field $\phi$ (the "Higgs mass") \be
 \xi=1/m_{\phi},~~m^2_{\phi}=2\lambda \phi^2_0 \ee
 As it is known [23] the London's limit for the superconductor of the
 second kind, corresponds
 to the relations
 \be
 \delta\gg\xi, ~~или~~ e\ll \lambda \ee
One can also calculate the string energy per unit length--the
string tension -- for the string of the minimal magnetic flux $2\pi$.
Using (41) and (47-48) one obtains [23]
\be
 \sigma_{ANO} =
\frac{\pi}{\delta^2} ln\frac{\delta}{\xi},~~ \delta/\xi \gg 1 \ee
 Note that the main contribution to
  $\sigma_{ANO}$ (50) comes from the term
$|\vec D \phi|^2$  in (41).

Thus the physical picture of the ANO string in  hte London's limit implies the mass generation
of the magnetic quanta, $m=1/
\delta$,
 which are much less than the Higgs mass
$m_{\phi}=1/\xi$.

It is instructive to see how the screening of the magnetic field
(mass generation) occurs:

First magnetic  field creates around its flux (field--strength line)
the circle of the current $\vec j$ from the superconducting medium,
as described by London's  equations (43). Then Maxwell equation (42)
tells that around the induced current $\vec{j}$ there appears a
circulating magnetic field, directed \underline{opposite} to the
original magnetic field $\vec{B}$ and proportional to it.
 As a result magnetic field is partly screened in the
middle and finally completely screened far from the center of the
magnetic flux (which is the ANO string).

The profile of the string, i.e. $B_{11}(r)$ as a function of the
distance of the string axis is shown in Fig.6, as obtained from Eqs.
(47).

It is interesting to compare this ANO--string profile with the
corresponding profile obtained in the gluodynamics. As was discussed
in the previous chapter, the distribution of the parallel component
of the  colorelectric field  was measured in [24]  and
is  also exponentially decreasing far off
axis, as can be seen in Fig.6, and both profiles of the ANO string
 and  of the QCD string in Fig.6 are very much similar, thus
supporting the idea of dual Meissner mechanism .

To look more closely at the similarity of classical equations
(4.9-4.10) with the corresponding equations, obtained in lattice
Monte--Carlo simulations, one needs first an  instrument to recognize
the  similarity of effective Lagrangians in the model
(40) and in  QCD. This is  discussed in  the
next chapter.

We conclude point 2) of this chapter discussing
parameters
 $\delta_{QCD}$ and   $\xi_{QCD}$,
parameters dual--analogical to $\delta$ and $\xi$ of (44-45).
This is done via the Abelian projection method  in [25] with the result
(see Fig. 7)
 \be
 \delta_{QCD}\sim \xi_{QCD}\sim 0.2 fm
 \ee

 Thus in QCD
situation is  somewhat in the middle between the type I and type II
dual superconductor.
 The calculation of the effective potential
$V(\phi)$  in SU(2) gluodynamics, made recently in [26],
shows a two-well structure, but with a rather shallow well, hence the effective
$\lambda_{QCD}$  in the Lagrangian (40) is not large, again in agreement with
(51). A very interesting discussion of properties of dual monopoles and their measurement
on the lattice is contained in [27].

3) Condensate formation and symmetry breaking\\

The phenomenon of
 superconductivity is usually associated with the formation
of the condensate of Cooper pairs  (although it is not necessary).

The  notion of condensate is most clearly understood for the
noninteracting Bose--Einstein gas at almost zero temperature, where
the phenomenon of the Bose-- Einstein condensation takes place.
Ideally in quantum--mechanical systems condensate can be considered
as a coherent state .

When interaction  is taken into account, the meaning of the
condensate is less clear [28].

In quantum  field theory one associates condensate with the
properties of the wave  functional and/or Fock columns. Again in the
case of no interaction one  can construct the coherent state in the
second--quantized formalism (see e.g. the definition of wave operators
in the superfluidity case in [29])

One of the properties of such a state is the fixed phase of the wave
functional, which means that the $U(1) $ symmetry is violated.

The simplest example is given by the Ginzburg--Landau theory (40),
where in the approximation when $\lambda\to \infty$, the  wave
functional $\Psi\{\varphi\}$ can be approximated by the classical
solution $\varphi =\varphi_0$, $\Psi\{\varphi
\}\to\Psi\{\varphi_0\}$.

The solution $\varphi=\varphi_0$, where $\varphi_0$ has a fixed phase,
violates U(1) symmetry of the Lagrangian (40), and one has the
phenomenon of spontaneous symmetry breaking (SSB) [30].
The easiest way to exemplify the SSB is the double--well Higgs
potential like in (40). Therefore if one looks for the dual
Meissner mechanism in QCD (gluodynamics) one may  identify the
magnetic monopole condensate $\tilde{\varphi}$, dual to the Cooper
pair condensate $\varphi=\varphi_0$ and find the effective potential
$V(\tilde{\varphi})$,  demonstrating that it has a typical
double--well shape. Such an analysis was performed in [26] using the
Abelian projection method and will be discussed in the next chapter.

4) Finally in this  chapter we shall look at the dual Meissner
mechanism from the point of view of field and current correlators [31].
This will enable us to formulate the mechanism in the most general
form, valid both in (quasi) classical and quantum vacuum.

One can use the correlator (16) to study confinement of both
magnetic and electric charges (the latter was discussed in chapter
3 -- Eq.  (19) and subsequent text). Let us rewrite (16) for
correlators of electric and magnetic  fields separately \be
<E_i(x)E_j(y)>=
\delta_{ij}(D^E+D^E_1+h^2_4\frac{\partial D_1^E}{\partial
h^2})+h_ih_j\frac{\partial d^E_1}{\partial h^2}
\ee
\be
<H_i(x)H_j(y)>=
\delta_{ij}(D^H+D^H_1+h^2\frac{\partial D_1^H}{\partial
h^2})-h_ih_j\frac{\partial d^H_1}{\partial h^2}
\ee
where $h_{\mu}=x_{\mu}-y_{\mu},~~h^2=h_{\mu} h_{\mu}$.

In (52)-(53) we have specified correlators $D,D_1$ for electric
and magnetic fields separately, since in Lorentz--invariant vacuum
$D^E=D^H$, $D^E_1=D^H_1$, otherwise, e.g. in the Ginzburg--Landau
model, electric and magnetic correlators may differ, as also in any
theory for nonzero temperature .

Now let us compare the Wilson loop averages for electric and magnetic
charges. In case of electric charges the result is the area law
(17) with the function $D\to D^E$ responsible for confinement.

Now consider a magnetic charge in the contour C in the 14 plane, the
corresponding Wilson loop is
\be
<\tilde{W}(C)>=<exp(ig\int\tilde{F}_{14}d\sigma_{14})>= exp
(-\sigma^*S_{min})
\ee
Here $\tilde{F}_{\mu\nu}$ is the dual field,
$\tilde{F}_{\mu\nu}=\frac{1}{2}
e_{\mu\nu\alpha\beta}F_{\alpha\beta}$, and $\tilde{F}_{14}=H_1$. From
(17) one obtains
\be
\sigma^* =\frac{g^2}{2}\int d^2xD^H_1(x)(1+O(T_g^2H))
\ee
Eq. (55) is kind of surprise. For electric charges $D^E$ yields
confinement and is nonperturbative, supported by magnetic monopoles,
cf. Eq. (22), while $D^E_1$ contributes perimeter correction to
the area law and contains also perturbative contributions like
Coulomb term.

The same electric charges in the (23) -- plane Wilson loop bring
about again the area law (17) with $D\to D^H$. Also $D^H$ is
coupled to magnetic monopoles in the vacuum, indeed taking divergence
from both sides of (53) one gets \be <div H(x), div
H(y)>=-\partial^2G^H(x-y) \ee Instead confinement of external
magnetic charges, Eq. (4.16), is coupled to the function $D_1^H$ (or
$D_1^E$ if one  takes $<\tilde{W}(C)>$ for the loop in the (23)
--plane).

Thus duality of electric -- magnetic  extemal charges requires
interchange $D^H,D^E\leftrightarrow D_1^E,D_1^H$.

At this point one should be careful and separate out perturbative
interaction which is contained in $D_1$, namely one should replace in
(55) $D^H_1$ by $\tilde{D}_1^H$ where
\be
\tilde{D}^H_1=D_1^H-\frac{4e^2}{x^4}
\ee
It is important to stress again that it is the nonperturbative
contents of correlators which may create a new mass parameter like
$\sigma^*$ and which should enter therefore in correlators (the
perturbative term contributes the usual Coulomb--like interaction
which is technically easier consider not as a part of $D_1^H$, but to
separate at earlier stage, see e.g. [32]).

This mass creation can be visualized in the Ginzburg--Landau model,
where $D_1$ can be computed explicitly from (40)
\be
D_1^{LG}(x-y)=(e^2|\phi|^2-\partial^2)^{-1}_{xy}\approx e^{-m|x-y|}
\ee
where in the asymptotic region $|\phi|=|\phi_0|$ and
$m=e|\phi|_0=1/\delta$.

From the point of view of duality $D_1^H(x)$ Eq.(49) should be
compared with the behaviour of $D^E(x)$, which was discussed above in
Section 3 and measured in [13].
\be
D(x)=D^E(x)=e^{-\mu|x|},~~\mu\approx 1 GeV
\ee
Eqs. (58-59) demonstrate validity of dual Meissner mechanism on
the level of field correlators.

\section{Abelian projection method}

In the previous section it was shown how the string is formed between
magnetic sources in the vacuum described by the Abelian Higgs model.

In the dual Meissner mechanism of confinement [3] it is assumed that
the condensate of magnetic monopoles occurs in QCD, which creates
strings between color electric charges. There are two possible ways
to proceed from this point. In the first one may assume the form of
Abelian or nonabelian Higgs model for dual gluonic field and scalar
field of magnetic monopoles. This type of approach was pursued in
[33] and  phenomenologically is quite successful: the linear
confinement appears naturally and even spin--dependent forces are
predicted in reasonable agreement with experiment [34]. We shall not
go into details of this very interesting approach, referring the
reader to the cited papers, since here lacks the
most fundamental part of the problem -- the derivation of this dual
Meissner model from the first principles -- the QCD Lagrangian.

Instead we turn to another direction which was pursued very
intensively the last 8-10 years -- the Abelian projection method
dating from the seminal paper by 'tHooft [4].

The main problem --how to recognize configurations with properties of
magnetic monopoles, which are responsible for confinement. The
'tHooft's suggestion [4] is to chose a specific  gauge, where
monopole degrees of freedom hidden in a given configuration become
evident.  The corresponding procedure was elaborated both in
continuum and on lattice [35] and most subsequent efforts have been
devoted to practical separation (abelian projection) of lattice
configurations and study of the  separated degrees of freedom, and
construction of the effective Lagrangian for them. We start with the
formal procedure in continuum for SU(N) gluodinamics, following
[4,35].

For any composite field $X$ transforming as an adjoint
representation, like $F_{\mu\nu}$, e.g.
\be
X\to X'=VXV^{-1}
\ee
let us find  the specific unitary matrix $V$ (the gauge), where $X$
is diagonal
\be
X'=VXV^{-1}=diag (\lambda_1,\lambda_2,...\lambda_N)
\ee

For $X$ from the Lie  algebra of  SU(N), one can choose
$\lambda_1\leq\lambda_2\leq\lambda_3\leq...\lambda_N$. It is clear
that $V$ is determined up to the  left multiplication by a diagonal
SU(N) matrix.

This matrix belongs to the Cartan or largest Abelian subgroup of
SU(N), $V(1)^{N-1}\subset SU(N)$.

Now we transform $A_{\mu}$ to the gauge (61)
\be
\tilde A_{\mu}=V(A_{\mu}+\frac{i}{g}\partial_{\mu})V^{-1}
\ee
and consider how components of $\tilde A_{\mu}$ transform under
$U(1)^{N-1}$. The diagonal ones
\be
a^i_{\mu}\equiv (\tilde A_{\mu})_{ii}
\ee
transform as "photons":
\be
a_{\mu}^i\to a^i_{\mu}=a^i_{\mu}+\frac{1}{g}\partial_{\mu}\alpha_i
\ee
while nondiagonal, $c_{\mu}^{ij}  \equiv (A_{\mu}^{ij}$, transform as
charged fields.
\be
C^{'ij}_{\mu}= exp [i(\alpha_i-\alpha_j)] C_{\mu}^{ij}
\ee

But as 'tHooft remarks [4], this is not the whole story -- there
appear singularities due to a possible  coincidence of two or more
eigenvalues $\lambda_i$, and those bear properties of magnetic
monopoles. To make it explicit consider as in [35] the "photon"
field strength $$ f^i_{\mu\nu}=\partial_{\mu}
a^i_{\nu}-\partial_{\nu}a^i_{\mu}= $$ \be
 =VF_{\mu\nu}V^{-1}+ig[V(A_{\mu}+\frac{i}{g}\partial_{\mu})V^{-1},~~
 V(A_{\nu}+\frac{i}{g}\partial_{\nu})V^{-1}]
 \ee
 and define the monopole current
 \be
 K^i_{\mu}=\frac{1}{8\pi}\varepsilon_{\mu\nu\sigma\rho}\partial_{\nu}
 f^i_{\rho\sigma},~~
 \partial_{\mu}K^i_{\mu}=0.
 \ee
 Since $F_{\mu\nu}$ is regular, the only singularity giving
 rise to $K^i_{\mu}$ is the commutator term in (66), otherwise the
 smooth part of $a^i_{\mu}$ does not contribute to $K^i_{\mu}$
 because of  the antisymmetric tensor.

 Hence one can define the magnetic charge $m^i(\Omega)$ in the 3d
 region $\Omega$,
 \be
 m^i(\Omega)=\int_{\Omega}d^3\sigma_{\mu}K^i_{\mu}=
 \frac{1}{8\pi}\int_{\partial\Omega}
 d^2\sigma_{\mu\nu}\varepsilon_{\mu\nu\rho\sigma}f^i_{\rho\sigma}
 \ee

 Consider now the situation when two eigenvalues of (61) coincide,
 e.g. $\lambda_1=\lambda_2$. This may happem at one  3d point in
 $\Omega$ , $x^{(1)}$, i.e. on the line in the 4d, which one can
 visualize   as the magnetic monopole world line. The contribution to
 $m^i(\Omega)$ comes  only from the infinitesimal neighborhood
 $B_{\varepsilon}$ of $x^{(1)}$
 $$
 m^i(B_{\varepsilon}(x^{(1)}))=
 \frac{i}{4\pi}\int_{S_{\varepsilon}}
 d^2\sigma_{\mu\nu}\varepsilon_{\mu\nu\rho\sigma}
 [V\partial_{\rho}V^{-1},V\partial_{\sigma}V^{-1}]_{ii}=
   $$
   \be
   =-\frac{i}{4\pi}\int
   d^2\sigma_{\mu\nu}\varepsilon_{\mu\nu\rho\sigma}\partial_{\rho}
   [V\partial_{\sigma}V^{_1}]_{ii}
   \ee
   The  term $V\partial_{\sigma}V^{-1}$ is singular and should be
   treated with care. To make it explicit, one can write.
   \be
   V=W\left(\begin{array}{ll}
   cos\frac{1}{2}\theta+i\vec{\sigma}\vec e_{\phi}
   sin\frac{\theta}{2},& 0\\
   0,&1
   \end{array}
   \right)
   \ee
   where $W$ is a smooth SU(N) function near $x^{(1)}$.
   Inserting it in
  (69) one obtains
\be
m^i(B_{\varepsilon}(x^{(1)}))=\frac{1}{8\pi}
\int_{S_{\varepsilon}}d^2\sigma_{\mu\nu}
\varepsilon_{\mu\nu\rho\sigma}\partial_{\rho}
(1-cos \theta)\partial_{\sigma}\phi[\sigma_3]_{ii}
\ee
where $\phi$ and $\theta$ are azimuthal and polar anges.

The integrand in (71) is a Jacobian displaying a mapping from
$S^2_{\varepsilon}(x^{(1)})$ to $(\theta,\phi)\sim SU(2)/U(1) $.

Since
\be
\Pi_2(SU(2)/U(1))=Z
\ee
the magnetic charge is $m^i=0, \pm 1/2, \pm 1,..$.

From the derivation above it is clear, that
the point
 $x=x^{(1)}$, where
 $\lambda_1(x^{(1)})=\lambda_2(x^{(1)}),$ is a singular
 point of the gauge transformed
 $\tilde A_{\mu}$ and $a^i_{\mu}$, and the latter behaves near
 $x=x^{(1)}$ as $0(\frac{1}{|x-x^{(1)}|})$,
 and abelian projected field strength
 $f^i_{\mu\nu}$ is
 $0(\frac{1}{|x-x^{(1)}|2})$,
 like the field of a point--like magnetic monopole.

 However several points should be stressed now

 i) the original vector potential
  $A_{\mu}$  and
 $F_{\mu\nu}$  are smooth and do not show any singular behaviour

  ii) at large distances
 $f^i_{\mu\nu}$  is not, generally speaking, monopole--like, i.e. does not  decrease
  as
  $\frac{1}{|x-x^{(1)}|2}$,
  So that similarity to the magnetic monopole can be seen only in topological properties
  in the vicitity of the singular point
  $x^{(1)}$.

  iii) the fields $A_{\mu},a_{\mu}^i$  have in general
  nothing to do with classical
  solutions, and may be quantum
  fluctuations. Actually almost any field distribution in the vacuum
  may be abelian projected into $a_{\mu}^i,f_{\mu\nu}^i$  and
  magnetic monopoles then can be detected.

  Examples of this  statement will be given below, but before doing that we must say
  several words about the choice of the field
  $X$ in (60) and more generally about the explicit gauge choice.

  By now the most popular choices for the adjoint operator
   $X$, (60) is for nonzero temperature the fundamental
   Polyakov line
  \be
  L_{ij}(\vec x)=(P exp ig\int^{\beta}_0 dx_4 A_4(x_4,\vec x))_{ij}
  \ee
  where
  $\beta=1/T$, and $T$  is the temperature,
   $A_{\mu}(x_4,\vec x)$  is required to be periodic in
   $x_4$ and  $i,j$  are fundamental color indices.

   Another  widely used gauge is   the so-called maximal abelian  gauge
   (MAG) [35], which in lattice notations can be expressed as a gauge
   where the following quantity is maximized
   \be
    R
   =\sum_{s,\mu} Tr (\sigma_3\tilde U_{\mu}(s)\sigma_3\tilde
   U^+_{\mu}(s)) \ee Here the link matrix $U_{\mu}\sim exp ig
  A_{\mu}^{(s)}\Delta z_{\mu}$,  and $\tilde U$ is gauge transformed
  with the help of $V$ \be \tilde
  U_{\mu}(s)=V(s)U_{\mu}(s)V^{-1}(s+\mu) \ee
  In the continuum  MAG is characterized by the condition, which in
  the SU(2) case looks most simply \be (\partial_{\mu}\mp ig
  A^3_{\mu})A_{\mu}^{\pm}=0,~~ A_{\mu}^{\pm}=A^1_{\mu}\pm iA^2_{\mu}
  \ee
  As will be seen the choice of gauge in the abelian projection method is crucial:
  e.g. for the minimal  abelian gauge, corresponding to the minimum of
  $R$ (74), Abelian projected monopoles have no influence on
  confinement [36].

  Since the total  of papers on abelian projection is now
  enormous, let us discuss shortly the main ideas and results. The
  most part of results are obtained doing abelian projection on the
  lattice. (A short introduction and discussion of lattice technic is
  given in [27]). The separation of monopole degrees of freedom was
  done as follows.  For an abelian projected link (75) one can
  define the $U(1)$  angle $\theta_{\mu}$ \be \tilde U_{\mu}=exp
  (i\theta_{\mu}\sigma_3),~~-\pi\leq\theta_{\mu}\leq\pi
  \ee
  Then for the plaquette $\tilde U_{\mu\nu}\sim \tilde U_{\mu}\tilde
  U_{\tilde{\nu}}\sum exp (i\theta_{\mu\nu}\sigma_3)$
  one can write  -- $-4\pi\leq \theta_{\mu\nu}\leq4\pi$    and
  define the "Coulomb part" of
  $\theta_{\mu\nu},\bar{\theta}_{\mu\nu}=mod_{2\pi}\{\theta_{\mu\nu}\} $,
  so that
  \be
  \bar {\theta}_{\mu\nu}=\theta_{\mu\nu}+2\pi n_{\mu\nu}
  \ee
  and  $n_{\mu\nu}$ counts the number of Dirac strings across the
  plaquette $(\mu\nu)$.

  Now one can calculate different observables in AP and find the contributuion
  to them separately of the "Coulomb part"
  $\bar{\theta}_{\mu\nu}$  and the monopole part
   $m_{\mu\nu}\cong
  2\pi n_{\mu\nu}$. This was done in a series of papers and the monopole dominance was demonstrated
  i) for the string tension [38]
  ii) for fermion propagators and hadron masses [39]
  iii) for the topological susceptibility [40]
  . In Fig.8 a comparison is made of the total AP
  contribution to the string tension and of the monopole part, which is seen
  to dominate as compared to the Coulomb part.

  Another line of activity in AP is the derivation of effective
  Lagrangians for AP degrees of freedom. ( See e.g. in [41]). The
  resulting Lagrangian however  does not confine AP neutral objects
  like "photons" and  the latter should contribute to the spectrum in
  contradiction to the experiment.
            Therefore the Lagrangian  is not very useful both
  phenomenologically and fundamentally. We shall not discuss these
  Lagrangians referring the reader to the cited literature.
  Let us come back to the investigation of the confinement mechanism
   with  the help of the AP method.

     A direct check of the dual
   Meissner mechanism of confinement should contain at least two
   elements:  a detection of dual London current and a check  of
   magnetic monopole condensation.  The first was done in [42]. The
   dual of the London equation (43) is \be \vec E=\delta^2
    rot j^{\to}_{M},~~\delta=1/m
     \ee
      For the ideal type -II
    superconducting picture one needs that $\delta\gg \xi$, where
  $\xi =m^{-1}_{\phi}$ is  the Higgs mass (corresponding to the
   magnetic monopole condensation). In practice in [42] it was found
  that $\delta\sim\xi$, and therefore the condensate is "soft",
  implying that exact solutions of ANO string are necessary to
  compare with.  Such analysis was done in [43] for $SU(2)$  and
  $SU(3)$  and recently in [44].

   The string profile
   $\rho(x_{\bot})$  was also studied using AP,
   and the resulting
  $\rho(x_\bot)$ [45] is similar to the one, obtained in the full lattice  simulation
  [46].
  Thus the whole picture of currents and density is compatible with
  the dual superconductivity.

Now we come to the second  check -- of monopole condensation. Of
special interest is the problem of definition of monopoles. In the
first lattice studies [35] the AP monopoles in the maximally Abelian
gauge have been indentified through magnetic currents on the links of
the dual lattice, and the perimeter density of the currents was
measured below and above transition temperature $T_c$, showing a
strong decrease of this density  at $T>T_c$.

Later it was realized [47] that monopole density defined in this way
may not be a good characteristics of dual superconductivity and
monopole condensate, and for the latter one needs the creation
operator of the magnetic monopole. In this way one can define the
dual analog of the Higgs field $\varphi$, and settle the question of
condensation. The general mathematical construction of the monopole
creation operator was given before in [48], and several papers
[49] used this  U(1) construction for the AP monopoles.

Another construction was used for the U(1) theory in  [50] and as in
[49] the condensation of monopoles was also demonstrated. In the
SU(2) case the analysis was performed in [51] and [26].

In the last paper the important step was done in finding an effective
potential $V(\varphi)$  for the monopole cration operator $\varphi$
(defined similarly to Froelich--Marchetti [48], the exact equivalence
to [48] is still not proved in [26]).

If one believes in the dual Meissner picture of confinement,  then
one should expect the two--well structure of $V(\varphi)$, symmetric
with respect to the change
 $\varphi \to -\varphi$.  One can see in Fig.9a the r.h.s. of
  $V(\phi)$ (for positive  $\phi$)
   which indeed has a minimum at
    $\phi = \phi_c$,
   shifted to the right  from
    $\phi =
     0$, for values of  $\beta = \frac{4}{g^2}$  in the region of
     confinement.
     In  Fig.9b the same quantity
      $V(\phi)$ is measured in the deconfinement phase, and as
     one can see, the minimum  $\varphi=\varphi_c$ is at zero,
      $\phi_c =0$. In this way the analysis of [26] gives an evidence
     of AP monopole condensation.
      Note, however, that strictly speaking the condensation
     should be proved in the  London limit
     $(\lambda \to \infty)$, otherwise quantum fluctuations of
      $\varphi$ might prevail for a shallow well like that in Fig.9a
      (we remind that the system has a finite number of degrees of
     freedom in a finite lattice volume).
     Till now we have said nothing about the nature of configurations
     which due to AP disclose the magnetic monopole structure
     and   ensure confinement. They could be classical
     configurations or quantum fluctuations (the last
     possibility is preferred by the majority of researches).

     Recently  an interesting AP analysis of classical
     configurations   has been done [52-54]. We shall mostly pay
     attention to the first paper [52], created a wave of
     further activity. The authors of [52] make AP analytically
     of an isolated instanton, multiinstanton and
     Prasad--Sommerfield monopole and demonstrate in all cases
     the appearance of a straight--line monopole current. In the
     first case the current is concentrated at the centre of
     instanton and its direction depends on the parametrization
     chosen. Later on in [53] the same type of analysis was made
     numerically with the instanton--antiinstanton gas, and  the
     appearence of monopole--current loops was demonstrated, of
     the size of instanton radius and stable with respect to
     quantum fluctuations.

     Hence everything looks as if there is a hidden magnetic
     monopole inside an instanton. This result is extremely
     surprising from several points of view. First of all the
     flux of magnetic monopole through the Wilson loop is equal
     to
     $\pi$ (this will be shown in the next Section), whereas the
     same flux of instanton is equal to zero (modulo
 $2\pi$), this fact helps to understand why the monopole gas may
     ensures  confinemet, while
      the instanton gas does not.  Therefore  the identification
     of monopoles and instantons is not possible. Secondly,
     confinement in the instanton gas was shown to be absent in
     several independent calculations [55],  and the appearance of
     rather large monopole loops in this gas [53] looks
     suspiciously. To understand  what happens in the AP method,
     and whether it can generate monopoles where they were originally
     absent, we turn again to the case of one instanton [52] and
     take into account that AP contains a singular gauge
     transformation, which transforms an originally smooth field
      $F_{\mu\nu}$  into a singular one, namely [52]
      \be F_{\mu\nu}(x)= V F_{\mu\nu}(x)
     V^{-1}+ F_{\mu\nu}^{sing},
      \ee
      where
       \be F_{\mu}^{sing}(x)= \frac{i}{g}
     V(x)[\partial_{\mu}\partial_{\nu}- \partial_{\nu}\partial_{\mu}]
     V^{-1}(x)
      \ee
      The matrix  $V(x)$  is singular and
     $F_{\mu\nu}^{sing}$  therefore does not vanish. Hence
     the correlator function
       $D(x)$,  of the fields (15) and currents (22)
     aquires due to this singular gauge transformation a
     new term
      \be D(x)\to D(x)+ D^{sing}(x), \ee
     where
      \be D^{sing}(x-y)\sim <F^{sing}(x) F^{sing}(y)> \ee
      The same
      $F^{sing}_{\mu\nu}$  causes the appearence of a
     magnetic monopole current which passes  through the center of an
     isolated instanton [52]
     .  Therefore the AP method indeed inserts a  singularity of
     magnetic monopole type into practically any configuration and
     therefore cannot be a reliable method of separation of confining
      configurations. On the other side the use of AP for
     lattice configurations reproduces well the confinement
     observables, e.g. the string tension [38], which means that
     confining contributions pass the AP test very well.

     From this point of view it is interesting to consider
     another example -- the dyonic (Prasad--Sommerfield)
     solution, also studied in [52]. It appears that the AP
     monopole current passes exactly  through the multiinstanton
     centers and coincides with the trajectory of the physical
     dyon, which can be calculated independently, i.e. the AP
     magnetic monopole current coincides with the total magnetic
     monopole current. This again supports the idea that the
     genuine confining configurations are well projected by the
     AP method.

 \section{Classical solutions as a possible source of confinement}

 In  the last Section we have seen, that the AP method gives no
 answer to the question what are the confining configurations,
 they  could be both classical fields and quantum fluctuations.
 Nothing about it can be said till now from the field
 correlators, however lattice measurements of correlators yield
 some information on the possible profile  of confining
 configurations.

 At the same time an intersting information can be obtained from
 the lattice calculations using the so--called cooling  method
 [56], where at each step of  cooling quantum fluctuations are
 suppressed more and more, and configurations evolve in the
 direction of decreasing of the action. Roughly these results
 demonstrate that out of tens of thousands of original
 configurations (which are mostly quantum noise) at some step of
 cooling only few (sometimes 15-25) are retained, which ensure
 the same string tension, as the original -- "hot " vacuum. With
 more cooling the number of configurations drops to several units
 (they are mostly (anti)instantons) and confinement disappears.

 Thus one can think that confining configurations in the vacuum
 differ from usual quantum fluctuations, and their action is
 probably larger than instantonic action or they are less stable.

 Therefore it is interesting to  look carefully into all existing
 classical solutions and check whether one can build up the
 confining vacuum out of those solutions.

 In this chapter we shall study several classical solutions:
 instantons, dyons and lattice periodic instantons, and give a
 short discussion of some other  objects, like
 torons.

  After disposing the individual properties of those we specifically
  concentrate on the contribution of each of the object to  the Wilson
  loop--what we call the elementary flux
  of the object -- and argue that flux proportional to  $\pi$ of
  dyons  and  twisted  instantons with $Q=1/2$ may yield
  confinement, in contrast to the case of instantons with flux  equal
  to 2$\pi$.

  To prove this one should construct the dilute gas of objects, which
  we do in the most nontrivial example of dyons.\\

  {\bf 6.1.} Classical solutions i.e. solutions of the equation
  \be
  D_{\mu}F_{\mu\nu}=0
  \ee
  can be written  in the (anti)selfdual case in the form of the
  so--called 'tHooft ansatz [57] or in the most general form of
  Atiya--Drinfeld--Hitchin--Manin (ADHM) [58].

  In the first (simpler and less general) case one has
  \be
  A_{\mu}^a=-\frac{1}{g}\bar{\eta}^a_{\mu\nu}\partial_{\nu}ln W,
  \ee
  where $\bar{\eta}^a_{\mu\nu}$ is the 'tHooft symbol,
  \be
  \bar{\eta}^a_{\mu\nu}=e_{a\mu\nu},\mu,\nu=1,2,3 ~~\mbox{and}
  ~~~\delta_{a\nu},\mu=4 ~~~ \mbox{or}~~~-\delta_{a\mu},\nu=4,
  \ee
  and $W$ satisfies equation $\partial^2W=0$, with a particular
solution
\be
W=1+\sum^N_{n=1}\frac{\rho^2_n}{(x-x^{(n)})^2}
\ee
Here $\rho_n, x_{\mu}^{(n)}, n=1,..N$ are real parameters. In the
simplest case, $N=1$, one has the instanton solution [59] with size
$\rho_n=\rho$ and position $x_{\mu}^{(1)}$. For $\rho_n$ finite
and $ N$ arbitrary (85),(87) gives a multiinstanton solution with
topological charge $Q=N$. In particular for $N\to \infty$,
$x_4^{(n)}=nb,~x_i^{(n)}=r_i$, one gets the Harrington--Shepard
periodic instanton [60], for which $A^a_{\mu}$ periodically depends
on (Euclidean) time $x_4\equiv t$.

A specific feature of instantons is their finite size: fields
$F_{\mu\nu}$ fall off at large distances from the center as
$x^{-4}$. This is very different from the case of magnetic
monopole, where fields decay as $x^{-2}$.

Another class of
solutions with these latter properties can be obtained from (85),
(87) in   the limit $\rho_n=\rho\to\infty$.

The case of $N=1$, when $W=\frac{1}{(x-x^{(0)})^2}$, yields no
solution, since it is a pure gauge.

The next case, $N=2$, is gauge equivalent to the (anti)instanton
with the position $\frac{1}{2}(x^{(1)}+x^{(2)})$ and size
$\frac{1}{2}|x^{(1)}-x^{(2)}|$.

We shall be interested in the case when $N\to \infty,
\rho_i=\rho\to\infty,$ $ x_4^{(n)}=nb,\vec x^{(n)}\equiv \vec{R}$ and
call this solution a dyon, since, as we demonstrate below, it has
both electric and magnetic long--distance field.

In this case $W$ can be  written as
\be
W\equiv \sum^{\infty}_{n=-\infty}\frac{1}{(\vec x-\vec
R)^2+(x_4-nb)^2},
\ee
and one can rewrite (88) using variables
\be
\gamma|\vec x- \vec R |\equiv r,~~ x_4\gamma\equiv t,~~
 \gamma=2\pi/b
\ee
as
\be
W=\frac{1}{2r}\frac{shr}{chr-cost}.
\ee

 From (85) one obtains vector--potentials
 \be
A_{ia}=e_{aik}n_k(\frac{1}{r}-cthr+
\frac{shr}{chr-cost})-\frac{\delta_{ia} sint}
{chr-cost},
\ee
 \be
A_{4a}=n_a(\frac{1}{r}-cthr+\frac{shr}{chr-cost}),
 \ee
 with $$\vec{n}=(\vec x-\vec R)/|\vec x-\vec R|.$$

  One can notice, that the dyonic field in this (singular, or
  'tHooft) gauge is now  long ranged in spatial coordinates
  \be
  A_{\mu a}\sim \frac{1}{r},~~ F_{\mu\nu}\sim \frac{1}{r^2}
  \ee
  and periodic in "time" $t$.

  Even more similarity to the magnetic monopole field can be seen,
  when one makes the (singular) gauge tranformation [61]
  \be
  \tilde A_{\mu}=U^+(A_{\mu}+\frac{i}{g}\partial_{\mu})U,~~
  U=\exp\biggl(\frac{i\vec{\tau}\vec{n}}{2}\theta\biggr),
  \ee
  with
  $$tg
  \theta=W_4\biggl[\frac{W}{r}+W_r\biggr]^{-1},
  ~~W_{\mu}\equiv\partial_{\mu}  W.$$
  In this gauge, sometimes called Rossi gauge, $\tilde A_{\mu}$
  looks like Prasad--Sommerfield solution [62]:
  \be
  \tilde A_{ia}= f(r)e_{iba}n_b,~~
  f(r)=\frac{1}{gr}(1-\frac{r}{shr}),
  \ee
  \be
  \tilde A_{4a}= \varphi(r)n_a,~~
  \varphi(r)=\frac{1}{gr}(rcth r-1).
  \ee

  Note that $\tilde A_{\mu a}$ does not depend on time, it
  describes a static dyonic solution, since it has both
  (color)electric and (color)magnetic field:
  \be
  E_{ka}=B_{ka}=\delta_{ak}(-f'-f/r)+n_an_k(f'-f/r+gf^2).
  \ee
  One may further gauge rotate $E_{ka},B_{ka}$ to the quasiabelian
  gauge,  where the only long--ranged component is along 3 axis:
  \be
  E'_{k3}=B'_{k3}(r\to\infty)\sim -\frac{1}{gr^2}n_k.
  \ee
  Eq.(98) justifies our use of the name dyon for the solution and
  demonstrates its clear similarity to the magnetic monopole. Note
  also, that $\tilde A_{4a}$ (96) is tending to a constant at
  spacial infinity, like the Higgs field component of the
  'tHooft--Polyakov monopole  [63].

  The total action of the dyon is calculated from (95),
  (96) or
  (91), (92) to be
  \be
  S=\frac{1}{2}\int d^3\vec r \int^T_0 dt
  (B^2_{ak}+E^2_{ak})=\frac{8\pi^2}{g^2b} T,
  \ee
  where $T$ is the length of the "dyonic string", in terms of the
  number $N$ of centers in (87) it is $T=b(N-1)$. For the given
  $N$ one also has
  \be
  S(N)=\frac{8\pi^2}{g^2} Q(N),~~ Q(N)=N-1.
  \ee

{\bf 6.2. }   This short section is devoted to another type of classical
    solutions -- those depending on boundary  conditions and
    defined in finite volume.
Here we consider torons and instantons on torus [64], which obey
    the twisted boundary conditions (b.c.) in the box $0\leq
    x_{\mu}\leq a_{\mu}$. Periodic b.c. are imposed modulo gauge
    transformation (twisted b.c.)
    \be
    A_{\lambda}(x_{\mu}=a_{\mu})=
    \Omega_{\mu}\biggl[A_{\lambda}
    (x_{\mu}=0)-i\frac{\partial}{\partial
    x_{\lambda}}\biggr]\Omega^+_{\mu}.
    \ee
    To ensure selfconsistency of $A_{\lambda}$ on the  lines, four
    functions $\Omega_{\mu} (\mu=1,..., 4)$
should satisfy conditions
\be
\Omega_1(x_2=a_2)
      \Omega_2(x_1=0)=\Omega_2(x_1=a_1)\Omega_1(x_2=0)Z_{12},
      \ee
      and analogous conditions for  $1,2\to i,j$, where
      $Z_{12}\epsilon Z(N)$ is the center of the group SU(N), \be
      Z_{\mu\nu}=exp \biggl(2\pi i
      \frac{n_{\mu\nu}}{N}\biggr),~~n_{\mu\nu}=-n_{\nu\mu}.
      \ee
      Here $n_{\mu\nu}$ are integers not depending on coordinates
      $x_{\mu}$.

      The twisted solutions $A_{\mu}$ (101) contribute to the
      topological charge
      \be
      \frac{g^2}{16\pi^2}\int_{|x_{\mu}|\leq
      a_{\mu}}Tr(F_{\mu\nu}\tilde{F}_{\mu\nu})d^4x=
      \nu-\frac{\chi}{N},
      \ee
      where $\nu$ -- an integer and $\chi=\frac{1}{4}
      n_{\mu\nu}\tilde{n}_{\mu\nu}=
      n_{12}n_{34}+n_{13}n_{42}+n_{14}n_{23}$.

       The action in the box is bounded from below
       \be
       \frac{1}{2}\int
       Tr(F_{\mu\nu}F_{\mu\nu})
       d^4x\geq\frac{8\pi^2}{g^2}|\nu-\frac{\chi}{N}|.
       \ee

        Consider e.g. the case of $n_{34}=-n_{12}=1$, all other
        $n_{\mu\nu}=0,\chi=1$. Then one has
        \be
        A_{\lambda}(x)=-\frac{\omega}{g}\sum_{\mu}
        \frac{\alpha_{\mu\lambda}x_{\mu}}{a_{\mu}a_{\lambda}},~~
        \alpha_{\mu\lambda}=-\alpha_{\lambda\mu},
        \ee
        where $\alpha_{12}=\frac{1}{2Nk};~~
        \alpha_{34}=\frac{1}{2Nl},~~ k+l=N,$
        \be
        \omega =2\pi\cdot diag(l,...,l,-k,...,-k),
                \ee
        the matrix $\omega$ has $k$ elements equal to $l$ and $l$
        elements equal to $(-k)$, and a condition is imposed
        $a_1a_2(a_3a_4)^{-1}=\frac{l}{k}=\frac{N-k}{k}$.
As a simple example take SU(2) and cubic box, then
        $k=l=1,\omega=2\pi\tau_3$ and
        \be
        A_{\lambda}(x)=-\frac{\tau_3}{a^2}
        \frac{\pi}{2g}\sum_{\mu}\bar{\alpha}_{\mu\lambda}
        x_{\mu},~~\bar{\alpha}_{12}=\bar{\alpha}_{34}=1.
        \ee
        This solution is selfdual  and the following relation
        holds
         \be Tr F_{\mu\nu}F_{\mu\nu}=Tr
         F_{\mu\nu}\tilde{F}_{\mu\nu}=
         \frac{16\pi^2}{\prod_{\mu}a_{\mu}N g^2}.
         \ee

          From (109) one can
           see that toron (108) is a particular
          case of a selfdual solution with constant field
          $\bar F_{\mu\nu}$
           \be
          A_{\mu}(x)=\bar{F}_{\mu\nu} x_{\nu}\frac{\tau_3}{2},
          \ee
          where the amplitude of constant field
          $\bar{F}_{\mu\nu}$ is quantized. For constant
          (anti)self\-dual field the analysis of Leutwyler [65]
          tells that such solutions are stable with respect to
          quantum fluctuations.

            The flux through the Wilson
          loop for the  solution (108) is the planes (12) or
          (34) is
          \be
          P exp (ig \int_C A_{\mu} dx_{\mu})=exp
          (-i\pi\frac{S}{a^2}\tau_3),
          \ee
          where $S$ is the area, bounded by the contour $C$.

          As we shall see in the next section the flux equal to
          $\pi$ for $S=a^2$ is a property very important for
          confinement. Another interesting property of torons,
          not shared by any other solutions, that its action is
          proportional to $1/N_cg^2$ and therefore stays constant
          for large $N_c$, where $g^2=g^2_0/N_c$. We come back to
          this property in conclusions.

          Another type of twisted solutions are twisted
          instantons [64]. These are  solutions with
          topological charge $Q$ (104) and $\nu$ integer non zero.

          Those solutions have been seen on the lattice [66], and
          the profile (distribution of $tr F^2_{\mu\nu}(x))$ is
          very close to that of the usual instanton.

          Unfortunately the analytic form of twisted instanton
          is still unknown; the top charge was found to be
          1/2 [66], and the  extrapolated string tension is
          probably nonzero.  These two
          facts are not accidental -- in the next section we show
          that halfinteger top. charge ensures a flux of $\pi$ and
          that in turn may lead to confinement.   \\

          {\bf    6.3.}  In this subsection we compute elementary flux
             inside Wilson loop for (multi)instanton, dyon and
          twisted instanton and connect properties of elementary flux
          to confinement in the gas of classical solutions [67].

Consider a circular Wilson loop in the
plane (12) and   take $A_{\mu}$ in
the
 form of the 'tHooft ansatz (5.2), where $N$ is fixed and
$x_i^{(n)}=0,x_4^{(n)}=nb$.  In this way one can study the case of
instanton $(N=1)$, periodic Harrington--Shepard instanton $(N\to
\infty, \rho_i=\rho$ fixed), multiinstanton $(N$ finite, $\rho_i$
finite) and dyon ($\rho_i=\rho\to \infty,~N\to\infty$).

When the radius of the loop  $R$ is much larger than the core of the
solution (i.e. $R\gg \rho$ for (multi)instantons or $R\gg b/2\pi$ for
dyon), the Wilson loop is \be W(C_R)=
exp\biggl(i\tau_3\pi  \frac{RW_r}{W}\biggr)\equiv exp(
i\tau_3\cdot flux ) ,
\ee where $W_r= \frac{\partial
W}{\partial|\vec{x}|}|_{|\vec{x}|=R},~|\vec{x}|\equiv r$.

 Now for (multi)instanton one  has for $R\gg \rho$
  \be
\frac{RW_r}{W}|_{r=R}=\frac{-\sum\frac{\rho_n^
2\cdot 2R}{(R^2+(x_4-nb)^2)^2}}{1+\sum\frac{\rho^2_n}
{R^2+(x_4-nb)^2}}\to 0.
\ee
In nonsingular gauge one  would  obtain for (multi)instanton the flux
$2\pi$ [68], in all gauges one has, that
\be
W(C_R)=1,~~(multi)instantons.
\ee
Consider now the case of dyons, which amounts to tending $\rho_n\to
\infty$ in $W_r$ in (112).

One can use the form (90) to obtain for dyon
\be
\frac{R W_r}{W}=-1;~~ flux =-\pi,~~ W(C_R)=-1.
\ee

It is amusing to consider also the intermediate  case of so--called
$\tau$--mo\-no\-poles [69], when $\rho_n\to\infty$, but $N$ is fixed,
so the length of the chain $L=Nb$ is finite. One can use (88) to
find two limiting cases:
  \be R\gg L,~~ \frac{RW_r}{W}=-2,~~flux
=-2\pi;~~ W(C_R)=1;
 \ee
  \be
   R\ll L,~~ \frac{RW_r}{W}=-1,~~flux
=-\pi;~~ W(C_R)=-1.
 \ee

Thus only $\tau$--monopoles long enough, i.e. almost dyons, may ensure
nontrivial Wilson loop, $W(C_R)\neq 1$.

To connect flux values (114)-(117)
 to confinement one can use model
consideration of  stochastic distribution of fluxes in the dilute
gas, as was done in [68,69]. More generally the picture of stochastic
fluxes was formulated in the model of stochastic confinement [70] and
checked on the lattice in [71].

We shall come back to the model of stochastic confinement in the next
Section, and now shall use simple arguments from [68,69].

Indeed, consider a
 thin 3d layer above and below the Wilson loop, of
thickness $l\ll R$, and assume that it is filled with gas of
(multi)instantons or dyons. If 3d density of the gas is $\nu$, so
that average number of objects  is $\bar{n}=\nu S l$, then Poisson
distribution gives probability of having $n$ objects  in the layer
around the plane of Wilson loop is
\be
w(n)=e^{-\bar{n}}\frac{(\bar{n})^n}{n!}.
\ee
If contribution to the Wilson loop of one object is
$\lambda,~\lambda=+1$ and $-1$ for instantons
 and dyons respectively,
then the total contribution is
\be
<W(C_R)>=\sum_ne^{-\bar{n}}\frac{\lambda^n\bar{n}^n}{n!}=e^{-\bar
n(1-\lambda)}= e^{-\sigma S},
\ee
where
\be
\sigma=(1-\lambda)\nu l.
\ee
Thus for instantons $(\lambda=1)$ one obtains zero string tension in
agreement with other calculations [55], while for dyons
$(\lambda=-1)$, confinement is present according to the model. Let us
stress, which features are important for  this conclusion of
confinement:\\ i) the flux =$\pi$, so that $W=-1$ for one  dyon,\\
ii) stochastic distribution of  fluxes, which enables us to use
Poisson (or \\similar) distribution,\\
iii) existence of finite thickness, i.e. of finite screening
length, so that objects more distant than $l$ completely screen each
other and do not contribute to the Wilson loop.

Notice that point iii) is necessary for area law, otherwise (for $l$
large, e.g. $l=R$) one obtains $\sigma $ growing with $R$,
i.e. superconfinement.

The same reasoning is applicable to torons and twisted instantons
[66].  Indeed from (111) one can see, that their elementary flux
is equal to ($-\pi$). Therefore if one divides all volume into a set
of twisted cubic cells, and ensures stochasticity of fluxes in the
cells one should have the same result (119) with
confinement present.

Torons [72] and twisted instantons [66] have been studied from the
point of view of confinement both analytically [72] and on the
lattice [66].  For torons the requirement  of stochasticity  is
difficult  to implement, since boundary conditions of adjacent cells
should ensure continuity of $A_{\mu}(x)$, and this introduces
ordering in the fluxes, and  confinement may be lost. In case of
twisted instantons with $Q=1/2$ [66] the field is essentially nonzero
around the center of instanton, and b.c. are much less essential.
The authors of [66] note a possibility of nonzero extrapolated value
for the string tension when the box size is increased beyond 1.2. fm.
It is still unclear what would be result when the twisted b.c. are
imposed only on the internal boundaries.\\

{\bf 6.4.} Below we concentrate on dyons, as most probable candidates for
classical confining configurations. One must study properties of dyon
gas and show that interaction in this gas is weak enough to ensure
validity of the dilute gas approximation. As it is usual, one assumes
the superposition ansatz
\be
A_{\mu}=\sum^{N_+}_{i=1}A_{\mu}^{+(i)}(x)+
\sum^{N_-}_{i=1}A_{\mu}^{-(i)}(x),
\ee
where $N_+,N_-$ are numbers of dyons and antidyons respectively. To
make the QCD vacuum $O(4)$ invariant, one should take any direction of
   the dyonic line, characterized by the unit vector
   $\omega^{(i)}_{\mu}$ and position vector $R^{(i)}_{\mu}$, so that
   \be
   A_{\mu}^{(i)}(x) =\Omega_i^+(LA)_{\mu}(r,t) \Omega_i.
   \ee
   Here $\Omega_i$ is the color orientation matrix and $L$ is the
   $O(4)$ (Lorentz) rotation matrix, while $r$ and $t$ are
   \be
   r=[(x-R^{(i)})^2-((x-R^{(i)})_{\mu}\omega_{\mu}^{(i)})^2]^{1/2},
   \ee
   \be
   t=(x-R^{(i)})_{\mu}\omega_{\mu}^{(i)}.
   \ee
   It is now nontrivial, in which gauge take solution $A_{\mu}$ in
   (122), e.g. one may take  singular gauge  solution (91) or
   time--independent one (95),(96). The sum (121) is not
   obtained by gauge transformation from one case to another. Indeed
   it appears, that the form (95),(96) is not suitable, since
   the action for the sum (121) in this case diverges  (see [67] for
   details).

   The form (91) is adoptable in this sense and we consider it in
   more detail.

    Since solutions fall off fast enough, cf. Eq. (93), the
   interaction between dyons, defined as $S_{int}$,

    \be
    S(A)=\sum_{i=1}^{N_++N_-} S_i(A^{(i)})+S_{int}
    \ee
    is Coulomb--like at large distances, e.g. for parallel dyon lines
   one has for two dyons
   \be
   S_{int}(R^{(1)},R^{(2)})=\frac{const\cdot
   T}{|\vec{R}^{(1)}-\vec{R}^{(2)}|},
   \ee
   where $T=Nb$ is the length of dyon lines. A similar
   estimate can be obtained for nonparallel lines.

   As was discussed in the previous section, the
   crucial point for the appearing of the  area law of
   Wilson loop is the phenomenon of screening. To
   check this, let us consider the field of tightly
   correlated pair $d\bar{d}$. When distance between
   $d$ and $\bar{d}$ is zero, the resulting
   vector potential is obtained using the superposition
   ansatz (121) and $d$ vector potentials
   (91)-(92 ) for $d$  and the corresponding one for
   $\bar d$, which differs from (91) by the sign of the
   last term and the total sign in (92):
    \be
   A_{ia}(d\bar{d})=\frac{2}{g}
   e_{aik}n_k(\frac{1}{r}-cthr+\frac{shr}{chr-cost}), \ee \be
   A_{4a}(d\bar d)=0.
   \ee
   At long range one has
   \be
   A_{ia}(d\bar d)=2e_{aik}\frac{1}{gr}+O(e^{-r}).
   \ee
   Calculation of $B_{ka}$  amounts to insertion in
   (98) $f=\frac{2}{gr}$, which immediately yields:
   \be
   B_{ia}(d\bar{d})=O(e^{-r}),~~E_{ia}\equiv O(e^{-r}).
   \ee
   Thus fields of $d$ and $\bar{d}$ completely screen
   each other at large distances; note, that this is
   purely nonabelian effect, since the cancellation is
   due to the quadratic term in $f$ in (98).

   Now take distance between $d$ and $\bar d$ equal to
   $\vec{\rho}=\vec{R}^{(1)}-\vec R^{(2)}$ and
   distance between observation point $\vec x$ and
   center of $d\bar d$ equal to $\vec{r}, ~~\vec
   r=\vec x - \frac{\vec R^{(1)}+\vec R^{(2)}}{2}$; assume
   that $r\gg \rho$. Then the field of $d\bar d$
   (averaged over direction of $\vec{\rho}$) is of the
   order
   \be
   B_k,E_k=O(\rho^2/r^4).
   \ee
   Hence contribution of the distant  correlated pair
   of  $d\bar d$ is unessential, and indeed in
   calculation of the Wilson loop one can take into
   account the distances to the plane of the loop
   smaller  than the  correlation length  $l$, which
   is  actually the screening length.

   From the dimensional arguments -- we have the only
   parameter in our Coulomb--like system -- the
   average distance  between the nearest
   neighbors $\nu^{-1/3}$, therefore one has
   \be
   l=c\nu^{-1/3},
   \ee
   where $c$ is some numerical constant, and $\nu$ is
   the 3d density of the dyon gas.

   Hence one expects the string tension in the dyonic
   gas to be of the order
   \be
   \sigma=c\nu^{2/3}.
   \ee
   Numerical calculations of $<W(C)>$ for the dyonic
   gas have been done in [73], but the density used was still
    much below that which is necessary for the  observation of
     the screening; work is now in progress.

     Summarizing this Section, let us discuss
     perspectives of the  classical solutions reported above
      as candidates for  confining configurations.
      Only two solutions, dyons and twisted  instantons, yield the suitable flux, equal
      to $\pi$, through the Wilson loop,
      therefore we discuss them separately.
      Dyons can be represented as a coherent chain of instantons of large  radius
       and correlated orientation of color field.
        When one goes from the instanton gas  to these coherent chains ,
        the action changes a little, but the enthropy decreases
        significantly, and possbly confinement occurs.
         To estimate the advantage or disadvantage of dyonic
         configurations from the point of view of the minimum of the
         vacuum free energy, one  must perform complicated
         computations, which are planned in the  nearest future.

         As to the twisted instantons, they require an internal
         lattice structure in the vacuum,  which may violate  the
         Lorentz invariance in some field
         correlators.

         After all, the problem  is solved, as in dyonic case, by the
         calculation of the free energy of the vacuum: in the nature
         there should come into existence that vacuum structure,
         which ensures the minimal free energy.  Lattice Monte-Carlo
          calculations satisfy this principle of minimal free energy
          (up to the finite size effects) and predict the confining
          vacuum with special nonperturbative configurations
          responsible for confinement.

          It is possible that those configurations are dyons, it is
          probable that they are not at all
          classical, but it is not excluded, that there exist unknown classical solutions,
          which ensure confinement after all.

   \section{Topology and stochasticity of
   confinement}
   In the previous chapter we have used the
   stochasticity of fluxes to obtain area law for
   dyons (magnetic monopoles). We start this chapter
   giving more rigorous treatment of this
   stochasticity  and comparing it to lattice data
   .

   For abelian theory magnetic flux through the loop
   $C$ is defined unambigously through the Wilson loop
   \be
   W(C)=exp~~ie\oint_C A_{\mu}dx_{\mu}=exp
   ~~ie\int_S\vec{H}d\vec{\sigma},
   \ee
   and the magnetic flux is
   \be
   \mu= e\int_S\vec{H}d\vec{\sigma}.
   \ee
   For SU($N$) theory
   the flux  can be  defined  analogously [70]  (we omit
    the word
     "magnetic", since it depends on the orientation of   the
   loop).

    Consider eigenvalues of the Wilson operator  (note absence of
   trace in its definition):
      \be
      U(C)=Pexp~~ ig \oint_C A_{\mu} dx_{\mu} \equiv \exp i
   \hat{\alpha}(C).
   \ee
   The eigenvalues of the unitary operator $U(C)$ are equal to $\exp
   i\hat{\alpha}(C)$, where $\hat{\alpha}(C)$ is  a diagonal matrix
   $N \times N$, depending on $A_{\mu}$.

   In electrodynamics $\alpha_n(C_{12})$ are
   additive for the contour
   $C_{12}$, consisting of two closed contours $C_1$ and $C_2$:
   \be
   \alpha_n(C_{12}) = \alpha_n(C_1) + \alpha_n(C_2).
   \ee
   In SU($N$) theory this is not so in general.
   Consider now the spectral density $\rho_c(\alpha)$, i.e.  averaged
   with the weight $\exp (-S_0 (A))$ the probability of the flux
   $\alpha(C)$:
   \be
   \rho_c(\alpha) = \frac{\int D A_{\mu}
   e^{-S_0(A)}\frac{1}{N}\sum^{N}_{m=1} \delta_{2\pi}(\alpha -
   \alpha_m(A_{\mu}, C))}{\int D A_{\mu} e^{-S(A)}},
   \ee
   where $S_0(A)$ is the standard action of the $SU(N_c)$ theory.

   Now any averaged Wilson operator over contour $C$, and also
   those for contour $C^n$, i.e. contour $C$, followed $n$ times, can
   be calculated with the help of $\rho_c(\alpha)$:
    \be <W(C^n)> =
   \int^{\pi}_{-\pi} d \alpha e^{in\alpha} \rho_c (\alpha).
    \ee
   Assume, that there is confinement in the system, i.e. the area
   law holds both for contour $C$ and $C^n$:
    \be <W(C^n)> = \exp
   (-k_n S).
    \ee

   Then the following equality holds [70,71]:
   \be
   \rho^C_S(\alpha) = \int^{\pi}_{-\pi} d \alpha_1 ...
   \int^{\pi}_{-\pi} d \alpha_n \rho^{C_1}_{S_1}(\alpha_1)
   ...\rho^{C_n}_{S_n} (\alpha_n)\delta_{2\pi}(\alpha - \alpha_1 -...
   -\alpha_n),
   \ee
   where the contour $C$ with the area $S$ is
    made of contours $C_i$
   splitting the area $S$ into pieces $S_i$.
   The proof [70,71] can be done in both directions: from
   (139),(140) to (141) and back. Sometimes this statement is
   formulated as a theorem [70]:  \underline{The necessary  and
     sufficient condition}\\
   \underline{
      of confinement is the additivity of
   random fluxes}.

   The randomness is seen in (141), which has the form of
   convolution as it should be for the product of probabilities for
   independent events.The additivity is evident from the argument of
   the $\delta$-function in (141).

   The density $\rho_c(\alpha)$ was measured in lattice calculations
   [71], and it was found, that $\rho_c(\alpha)$ indeed satisfies (141)
   and approximately  coincides with the $\rho_c^{d=2}(\alpha)$
   -- density for $d=2$ chromodynamics, if one renormalizes properly
   the charge. For the $d=2$ case $\rho_c(\alpha)$ is also known
   explicitly and satisfies (141) exactly [71].
   In that case confinement exists for trivial reasons.

   Let us now consider the nonabelian  Stokes theorem [11], which
   for the ope\-ra\-tor (136) looks like
   \be
   U(C) = P\exp ig \int_S d\sigma_{\mu\nu}(u) F_{\mu\nu}(u,x_0),
   \ee
   and take into account, that under gauge transformation $V(x)$ it
   transforms as
   \be
   U(C)\to V^+(x_0) U(C) V(x_0), \;\; F_{\mu\nu}(u,x_0) \to V^+(x_0)
   F_{\mu\nu} V(x_0).
   \ee
   Since $U(C)$ can be brought to the form $U(C) = \exp
   i\hat{\alpha}$ with $\hat{\alpha}$ diagonal by some unitary
   transformation, one can deduce, that this is  some gauge
   transformation $V(x_0)$, and, moreover, this is the same, which
   makes $\int d\sigma_{\mu\nu} F_{\mu\nu}(u, x_0)$
    diagonal.  Thus one can define the
    flux $\mu$
   similarly to (135):
   \be \hat{\mu} = diag \biggl\{ig V^+(x_0) \int_S d
   \sigma_{\mu\nu}(u) F_{\mu\nu}(u,x_0) V(x_0)\biggr\}.
    \ee
     Note, that
   dependence on $x_0$ presents in $U(C)$, cancels in $W(C) = tr
   U(C)$.

   The additivity of fluxes is seen in (144) explicitly.

   Now consider the statistical independence of fluxes, which obtain,
   when one divides the surface $S$ into pieces $S_1, ..., S_n$.
   Using the cluster expansion theorem [12] and discussion in Section
   3, one can conclude, that necessary and sufficient condition for
   this
   is the finite
   correlation length $T_g$, which appears in correlation functions
   (cumulants) $\ll F(1)...F(n)\gg$.  In this case, when pieces  $S_k,
   k=1,...,n,$ are all much larger in size than $T_g$, then different
   pieces become statistical independent. Thus our consideration in
   the framework of the field correlators in Section 3 is in clear
   agreement with the idea of stochastic confinement [70,71].
   The MVC in addition contains the quantitative method to
   calculate all observables in terms of given local correlators,
   which is absent in the stochastic
   confinement model [70,71].

    The appearance of  new physical quantity -- $T_g$ and cumulants
     is a further development of the   idea of stochastic
vacuum, which gives an exact quantitative characteristics of
randomness.

 When size of contours is of the order of $T_g$, the fluxes are no
more random, and   area law at such distances disappears -- there is
no area law at small distances, as is explained in Section 3.

 The
lattice measurements in [71] show that $\rho_c(\alpha)$ is strongly
peaked around $\alpha=0$ for small contours, when there is no area
law, which corresponds to the perturbative regime.

Instead for large contours the measured $\rho_c(\alpha)$ are rather
isotropic.

This  fact one can compare with our result  for  fluxes for instantons and
dyons. From our definition of fluxes (136) dyon in the plane of
contour $C$ corresponds to (cf.Eq.(112))
\be
\hat{\alpha}=\left(
\begin{array}{ll}
\pi&0\\
0&-\pi
\end{array}
\right),
\ee
and dyon with the center off the plane has smaller eigenvalues
$\alpha_m$. It is clear, that instantons with  flux zero cannot bring
about isotropic distribution of fluxes, while having maximal flux
$(\pm\pi)$ are most effective in creating the isotropic
$\rho_c(\alpha)$, when one integrates over all dyons in the layer
above and below the plane.

   It is instructive now to study the question of fluxes for adjoint
   Wilson loop and in general for Wilson loops of higher
   representations.

   One can keep the definition (136) also in this case, but
   $A_{\mu}$ and $\hat{\alpha}(C)$ should be expressed through
   generators of given representation:
    \be A_{\mu}=\sum_{a}A_{\mu
   a}T^a,~~ tr T^aT^b=\frac{1}{2}\delta_{ab}.
    \ee

Thus $\hat{\alpha}(C)$ for the Wilson loop in adjoint representation
is a matrix \\
$(N_c^2-1)\times(N_c^2-1)$. E.g. for SU(2) $(T^a)_{bc}=
\frac{i}{2}e_{abc}$.
To understand how stochastic wacuum model works for the adjoint
representation, let us take as an example the flux of one  dyon and calculate
$\hat{\alpha}_{adj}(C)$.
 Repeating our discussion,
preceding eq.(112) for large loops in  the (12) plane, one
concludes, that again only color index $a=3$ contribute, and one has
\be
\hat{\alpha}_{adj}(C)=\pi\cdot
 diag (T^3) lim \biggl(\frac{RW_r}{W}\biggr).
\ee
Since the last factor for dyon is $lim \frac{RW_r}{W}=-1$, and
$diag(T^3)=\left(
\begin{array}{lll}
1&&\\
&-1&\\
&&0\\
\end{array}
\right)$, one has finally
\be
\hat{\alpha}_{adj}(C)=\left(
\begin{array}{lll}
-\pi&&\\
&\pi&\\
&&0\\
\end{array}
\right).
\ee

Thus our conclusion of the elementary flux equal to
 $\pi$ holds true also in
the adjoint representation (and all higher representations), which gives
an argument for confinement of adjoint charges on the same ground as
for fundamental charges.

One has
\be
<W_{adj}>=\frac{1}{N^2_C-1} tr_{adj}e^{i\hat{\alpha}_{adj}(C)}=-1,
\ee
as well as $<\hat W_{fund}>=-1$.

So far we have  discussed stochasticity
 of the vacuum from the point
of view of fluxes and conclude, that it shows up as random
distribution of fluxes. In Section 3 the vacuum stochasticity was
formulated in the language of field correlators. Through the
AP method one can connect the latter with the
distribution of AP magnetic monopole currents.
(in the $U(1)$ theory an exact connection holds even without  AP).
One may
wonder,
 why magnetic monopoles or  dyons   are needed to  maintain the
stochastic picture of the vacuum?

To answer this question we start with the Abelian theory. Without
magnetic monopoles Bianchi identities $div \vec{H}=0$ are
operating, requiring, that all magnetic field strength lines are
closed.

This introduces strong ordering in the  distribution of magnetic
field, and no stochastic picture emerges.

As a result, confinement is not present in the system, as can be seen
from eq.(22). In presence of magnetic monopoles the  magnetic lines can
start and end at any place, where monopole is present, and one can
have a really stochastic distribution. As we discussed it in Section
3,
 the nonabelian  dy\-na\-mics can mimick  the effect of monopoles due
to triple correlators
$<E_iE_jB_k>$
 and ensure in this way the
stochastic distribution of fields.

So magnetic monopoles in Abelian theory, dyons in gluodynamics
create disorder in the system.

The same situation occurs in other spin and lattice systems, e.g. in
the  planar Heisenberg model the Berezinsky--Costerlitz--Thouless
vortices create disorder and and master the phase transition into the
high--temperature phase [74] (for details see also [1]).

All that is a manifestation of a general principle [2]:

\underline{Topologically nontrivial field configurations are responsible for
creation}\\
\underline{ of disorder and they drive the phase transition
order--disorder.}

From the QCD point of view the "ordered phase" is the perturbative
va\-cu\-um of QCD with long distance correlations $(D_1(x)\sim\frac{1}{x^4})$
and flux
distribution $\rho_c(\alpha)$
centered at zero, while the "disordered phase" is the real
QCD vacuum with short correlation length $T_g$ and with random
fluxes; there is no real phase transition in conti\-nuum: the phases coexist on two
different scales of distances (or moments). In the
lattice version of $U(1)$ theory there are indeed two phases: weak
coupling phase, corresponding to the usual QED, and strong coupling
phase with magnetic monopoles -- lattice artefacts -- driving the
phase transition.

The dyon is a continuum example of topologically nontrivial
configuration. In singular ('tHooft's)  gauge  dyon has
multiinstanton topological number, proportional to its length.

Dyon saturates the triple correlator
$<E_iE_jB_k>$
 and may be a source of
randomness of field distribution.

However the ultimate answer to the question about the nature of
confining configurations is still missing. The analysis of the dyonic
vacuum as a model of the QCD vacuum is not yet completed, and it is
possible that the topologically nontrivial configurations responsible
for confinement are dyons, or some other not known solutions or else
purely quantum fluctuations.

 We conclude this Section with the discussion of a possible
 connection between confinement and the Anderson localization [75].

 At the base of the similarity between these two phenomena lies the
 field stochasticity in the vacuum (medium), where quark (electron)
 propagates. The similarity however ends up just here.

 Namely, for an electron one can discuss its individual Green's
 function, which  always (for any density of defects) decays
 exponentially with distance. [There exists, however, a special
 correlator, e.g. the direct current conductivity $\sigma_{dc}$, which
 vanishes for localized states ( for large density of localized
 defects [76]) and is nonzero for  the delocalized states].

 In the case of a quark in the confining vacuum its Green's function
 (more precise: the gauge--invariant Green's function of the $q\bar
 q$ system averaged over vacuum configurations) corresponds to the
 linear potential, i.e. it behaves as
 $G(r)\sim exp (-r^{3/2})$, where $r$ is the distance between
  $q$ and $\bar q$. Thus the quark Green's function decays
 faster than any exponent, in contrast to the Green's function
 of an electron in the medium, always decaying exponentially.
 This property of vacuum Green's functions was coined by the
 author [77] the \underline{superlocalization}. If the average
 potential
 $\bar V$, acting on a quark, had been finite, then quark at some
 high energy could be freed and get to the detector.

  The essence of the superlocalization is exactly the fact, that the
 averaged potential
 $\bar V$ grows with distance without limits and therefore quarks are
 confined at any energy --this is the
 \underline{absolute confinement}.

 It is interesting to follow the mechanism, how the
 unbounded growth of
$\bar V$ occurs. To this end consider, as we did
 at the beginning of Section 3, a nonrelativistic quark, moving in
the
 $x,t$ plane, while the heavy antiquark is fixed at the origin.

 According to the quantum mechanical textbooks [78] the quark
Green's function is proportional to the phase integral, and using
the Fock--Schwinger gauge, one can write
 $$
G(X,T)=<exp~ig \int^T A_4(x,t) dt>\approx $$ \be \approx
1-\frac{g^2}{2}\int^T_0 dt\int^T_0dt'\int^X_0 du\int^X_0 du'
<E_1(u,t)E_1(u',t')>\approx \ee $$ \approx 1-\bar VT $$
Stochasticity of vacuum fields implies the finite
correlation length
  $T_g$  for the correlator
$<E_1(u,t)E_1(u',t')>$, i.e. according to (16) one has
 \be <E_1(u,t)E_1(u',t')>=D(u-u',t-t')+...  \ee  and for large
 $T$ and $X$ we obtain
   \be \bar V\approx const
|X|,~~|X|\to\infty \ee
Thus the linear growth of
 $\bar V$ is a consequence of random distribution of
\underline{field strength} $\vec E(u,t)$  and of the fact, that
 $\bar V$  is a result of the averaging of
\underline{vector-potentials} $A_{\mu}$, which are connected to
 $F_{\mu\nu}$ by an additional integral. This extra
integration causes the linear growth of
 $\bar V$,
and from the physical point of view this means the
 \underline{accumulation}  of fluctuations of the field
 $F_{\mu\nu}$  on the whole distance $X$ from the quark to the
antiquark.

This is the essence of the superlocalization phenomenon, which
still has no analogue in the physics of condensed matter.

\section{Conclusions}

In this rewiew we have looked at confinement from different
sides and described the mechanism of this phenomenon in the
language of field correlators, using more phenomenological
language of dual superconductivity -- effective classical
equations of the Ginzburg --Landau type, in the language of the
stochastic flux distributions and finally  we have studied
classical configurations which may be responsible for confinement.

All the way long we have stressed that confinement is the string
formation between color charges, the string  mostly consisting of
the longitudial colorelectric field.

 Let us try now to combine different descriptions of
confinement, given above in the review, and show in a simple
example how the string looks like.

To this end we use the  simple picture of the nonrelativistic
quark and the heavy  antiquark at the distance
 $X$  between them, discussed at the end of the last Section.
 From the  point of view of field correlators, confinement -- the
 string formation -- is the consequence of the fact, that  there
 exists the correlation length
  $T_g$, such that fields inside this length are coherent and those
 outside of this length are random. This is shown in Fig.10.1 (left
 part), where the strip of size
  $T_g$  is indicated in the plane (14) (one  could take instead any
 other
      plane, i.e. the plane (12) or (13)). Inside this strip the
 field is directed mostly in the same way (i.e. as on the string
 axis), but outside of the strip directions are random. The
 correlation length
  $T_g$  characterizes the string thickness (if the
 plane (12) or (14) is chosen). Thus
  $T_g$
  plays the double role, it gives the coherence lengh, where the
 string is created, and beyond which -- stochastic vacuum field,
 existing also before quark and antiquark have been inserted in the
 vacuum.

 Let us look at the same construction from the point of view of the
 dual superconductivity. Then one obtains the picture shown in
 Fig.10.2. Here an arrow depicts the monopole current
 $\tilde  j_{\mu}$,
which is caused by the colorelectric field
 $E_x$  of the string in accordance with the dual London's equation
 $rot\tilde j = m^2\vec E $.
The effect of this current is the squeezing of the string
field, which prevents the field flux lines from diverging into
the space, and as a result
 $E_x$
decays exponentially away from the string axis
 $Ox$ like $exp(-m\sqrt{y^2+z^2})$. Thus
 $m$  defines the string thickness and one can conclude
that  $m\sim 1/T_g$.
   And indeed Eq.(22) confirms this conclusion.

     Let us turn
now to the flux distribution and to the stochastic
vacuum model, Fig. 10.3. In this case the strip,
corresponding to the string in Fig. 10.1 can be divided
into pieces
 $S_1,S_2,...$  of the size  $d^2$, such that fluxes inside
 each piece are coherent and equal to e.g.
  $\pm \pi$  for the case of dyons, but two neighboring pieces are
 noncoherent -- their fluxes are mutually random. The string
 thickness is now built up due to the size
  $d$
 of the piece $S_n$, containing a coherent flux. If the surface
  $S_n$
 is penetrated by a monopole or a dyon, then
  $d$ coincides with the monopole or dyon size.

   To clarify this point let us find the minimal size of the loop
  $R$
 where the dyon flux is equal to  the asymptotic value of
  $(-\pi)$.  To this end we use (115) and insert there (90), and
 obtain the result, that for
 $R\gg
 b/2\pi\equiv \gamma^{-1}$ the flux is equal to
  $(-\pi)$ with exponential accuracy; hence the size
 of the dyon flux is equal to
 $\gamma^{-1}$ and this should be the size
 $d$ of the piece $S_n$.

 From the point of view  of field correlators
  $d$  should coincide with
 $T_g$, therefore the string thickness is of the order of the typical
 dyon (or  monopole) size (or an other classical solutions).

 Hence, all our pictures represented in Fig. 10.1- 10.3  can be
 combined in one generalized mechanism of the string formation,
 which is based on the  existence of coherent field domains of the
 size
  $T_g$, and beyond that size the fields are independent  and
  random.

  A question arises: who manages this structure of QCD vacuum and
  why in the case of QCD and gluodynamics the vacuum is made up
  this way, while in the case of QED and the Weinberg--Salam theory
  , the nonperturbative configurations are probably suppressed and
  the vacuum structure is different. To be able to answer this
  question is also to answer the question of phase transition
  mechanism for the temperature deconfinement, which was observed
  on the lattice [79]. This topic requires a separate review paper,
  since the amount of information accumulated here by now is very
  large. We shall confine ourselves to only few remaks on this point.

  Firstly, the density of the (nonperturbative) vacuum energy one
  can connect with the help of the scale anomaly theorem with the
  magnitude of the nonperturbative gluonic condensate [8]
 \be
 \varepsilon_{неперт}=
 +\frac{\beta(\alpha_s)}{16 \alpha_s}<F^a_{\mu\nu}(0)F^a_{\mu\nu}(0)>
 \ee
For small  $\alpha_s$  the function $\beta(\alpha_s)$
 is negative  --  in contrast to QED, and if it  keeps the sign in all
 effective region of
  $\alpha_s$,  then one can deduce that the nonperturbative vacuum
 shift (153) is advantageous, since it  diminishes the vacuum
 energy  ( and also the free energy for small temperatures).

 This conclusion can be considered as an intuitive idea why the
 nonperturbative vacuum in QCD is advantageous and comes into
 existance, while QCD -- not advantageous  and is not realized.

 Secondly, let us  briefly discuss the phase transition with
 increasing  temperature in QCD, referring the reader to lattice
 calculations [79] and original papers [81] for details, The main
 criterium which defines the vacuum structure preferred at a given
 temperature, is the criterium of the minimum of  the free energy
 (which is a corollary of the second law of thermodynamics). In the
 confining phase for
  $T>0$  the free energy consists of the term (153) and  of
  the contribution of hadronic excitations (glueballs,
  mesons and baryons), which slowly grows up to
   $T\sim 150
  MeV$.  Note that the gluonic condensate contains both colorelectric
  and colormagnetic fields, but only the first ones have to
  do with confinement in the proper sense of this term.

  The deconfinement phase, realized at
  $T>T_c$, usually was identified as the phase with perturbative
  vacuum, where quarks and gluons in the lowest order in
   $g$  are free [82]. However, from the point of view of the
  minimum of free energy it is advantageous to keep in the
  vacuum colormagnetic  fields and the corresponding  part of
  the condensate (153) since quarks in this case stay
  practically free, and one significantly  gains in energy
  -- around one half of the amount in (153).  This is the
  "magnetic confinement" phase [81]. Calculations in [81]
  yield
   $T_c$ in good agreement with lattice data for number of flavours
  $n_f=0,2,4.$

  The main prediction of the "magnetic confinement" is the
  area law for
  $T>T_c$  of the spacial Wilson loops [83], and the phenomenon of
  the "hadronic screening lengths", i.e. the existence of
  hadronic  spectra for
   $T>T_c$ in the Green's functions with evolution along space
  directions [81], which agrees well with the lattice measurements
  [79].

  Hence the picture of the phase transition [81] into the
  "magnetic confinement" phase, supported by computations, seems
  to be well founded. In this picture at
  $T>T_c$  disappear colorelectric correlators
  , more explicitly, the correlators of the type of
   of
  $D(x)$,  contributing to the string tension.

  What happens then with effective or real magnetic monopoles and
  dyons? In the AP method at
  $T>T_c$  the monopole density strongly decreases [35],
  which can be understood as an active annihilation or a
  close pairing of monopoles and antimonopoles. The same
  can be said about  pairs of dyons and antidyons. Thus
  the deconfinement phase of color charges can be
  associated with the confinement phase of monopoles (or
  dyons).
  However, the "magnetic confinement" phenomenon imposes definite
  requirements on the vacuum structure at
   $T>T_c$ .
  E.g.  there should exist magnetic monopole (dyonic)
  currents along the 4-th (euclidean time) axis --i.e.
  static or periodic monopoles (dyons). Such currents can
  confine in spacial planes (what is observed on lattices
  [83]), but do not participate in the usual confinement
  (i.e. in the temporal planes
  ($i4),i=1,2,3).$ These points will be elucidated in separate
  publications.

   We also had no space to discuss an important question of the
  connection between confinement and spontaneous  breaking of chiral
  symmetry and
   $U_A(1)$,
    where magnetic monopoles (dyons) can play an important role [84].

    We always held above that confinement is the property not only
  of QCD ( with quarks present in the vacuum), but also  of
  gluodynamics (without quarks). This conclusion follows from
  numerous  lattice data (see e.g. [10]), and also from computations
  supporting the dual Meissner effect as a basis of confinement
  discussed in the review, where quarks do not play an important
  role.

   On the other hand there are not any calculation or experimental
  data which support  a key role of quarks in the confinement
  mechanism.  For this reason we have not discussed above the model
  of V.N.Gribov (interesting by itself) and the reader is referred to
  original papers [85].

  The author is grateful to A.M.Badalyan for careful
  reading of the manuscript and many suggestions, to
  A.Di Giacomo and H.G.Dosch for interesting
  discussions and useful information, to
  M.I.Polikarpov, E.T.Akhmedov, M.N.Chernodub and
  F.V.Gubarev for useful and hot discussions of most
  points of this review, and to K.A.Ter-Martirosyan
  for constant support and discussions. The financial
  support of RFFR, grant 95-02-05436, and of INTAS,
  grant 93-79 is  gratefully acknowledged.

\newpage
\begin{center}
{\bf Figure captions }\\
\end{center}

Fig. 1.  Potential between static quarks in the triplet
representation of
 SU(3), computed in [10] on the lattice $32^4$. The solid line -- the
fit of the form
 $\frac{C}{R}+\sigma
R+ const.$ The potential and distance
$R$ are measured in lattice units
 $a$, equal to 0.055 fm for $\beta = 6.9$. Dynamical quarks are
absent.\\

 Fig. 2.
  The same potential as in  Fig. 1,  but with dynamical quarks
  taken into account in two versions (staggered fermions -- upper
  part,
  $a\simeq 0,11 $ fm, Wilson fermions -- lower part,
   $a\simeq 0,16 $ fm); calculations of Heller et al (second entry
  of [10]).\\

    Fig.3.
  The same potential as in  Fig. 1,  but  for quarks in the
  sextet (a) and octet (b) representation. Broken line -- the
  triplet potential of
   Fig. 1 multiplied by the ratio of Casimir operators, equal to
  2.25 for the octet and 2.5 for the sextet.\\

  Fig.4. Correlators   $D_{11}(x)=D+D_1+x^2\frac{\partial
  D_1}{\partial x^2}$ (lower set of points) and
 $D_{\bot}(x)=D+D_1$ (upper set of points) as functions of distance
 $x$. Crosses correspond to $\beta = 5.8,$  diamonds -- to
 $\beta =5.9$  and squares -- to $\beta =6.0$.  Solid lines are best
 fits in the form of independent exponents for
  $D(x)$  and $D_1(x)$.  Computations from [13].\\

     Fig. 5. Distribution of the parallel colorelectric
     field
     $E_{11}$ as a function of distance
     $x_{\bot}$ to the string axis. Measurements from [19] are made
     at different distances
      $x_{11}$ from the string end: filled circle -
     $x_{11}=3 a$, diamond --$x_{11}=5 a$ and triangle
      --$x_{11}=7 a$ ($a$ is the lattice unit ) .
     Solid line - Gaussian fit, dashed line -- calculation in [19]
     with the help of  $D(x)$  taken from  [13].\\

  Fig. 6. The same distribution  $E_{11}$,  as in
  Fig. 5, for SU(2) gluodynamics measured in [24] on the lattice
  24$^4$ for $\beta =2.7,
  ~~x_{11}=5a$, (the length of the string is  10a) -- empty circles,
  compared to the
  $B_{11}$ distribution in the Abrikosov string -- solid line. The
  growth od the solid  line at small $x_t$,  is unphysical and is due
  to violation of approximations made in (47).\\

 Fig. 7. Lattice measurements [25] of the penetration
 length
 $\delta\equiv \lambda$   and the coherence length
  $\xi $ as functions  $\beta=2N_c/g^2$  for AP configurations in
 SU(2) gluodynamics (upper two figures) and for SU(3) (the lower
 figure). The values of
  $\delta $  and
 $\xi$ are defined by comparison of field distribution and
 AP monopole currents with the solution of the
 Ginzburg--Landau equations.\\

Fig. 8.  The string tension measured in [43] for all AP
configurations  (squares) and separately for AP monopoles
(empty circles) and "photons" (filled circles) as functions
of
 $\beta=4/g^2$ in SU(2) gluodynamics.\\

               Fig. 9.  Effective potential of the AP
               monopole field
                $\varphi$, defined according to [48], measured in
               [26] for two values of
                $\beta$  in SU(2) gluodynamics, corresponding to
               confinement
               (Fig.9a) and deconfinement  (Fig.9b).\\

 Fig. 10.  The picture of string formation between a
 nonrelativistic quark and a heavy antiquark, is illustrated
 in three different approaches, discussed in the  text.\\

 (1) -- in the formalism of field correlators\\

 (2) -- in the formalism of dual superconductivity\\

 (3) --  in the picture of the stochastic flux distribution.

\end{document}